\documentclass[12pt]{article}
\usepackage[dvips]{graphicx}
\usepackage{epsf}
\usepackage{cite}
\usepackage{fancybox}
\usepackage{amsfonts}
\usepackage{amssymb}
\usepackage{amsmath}
\usepackage{bm}
\usepackage[dvips]{psfrag}

\makeatletter
\@addtoreset{equation}{section}
\makeatother
\begin{document}
%%%%%%%%%%%%%%%%%%%%%%%%%%%%%%%%%%%%%%%%%%%%%%%%%%%%%%%%%%%%%%%%%%%%%%%%%%%%%%%%%
%%%%%%%%%%%%%%%%%%%%%%%%%%%%%%%%%%%%% title %%%%%%%%%%%%%%%%%%%%%%%%%%%%%%%%%%%%%
%%%%%%%%%%%%%%%%%%%%%%%%%%%%%%%%%%%%%%%%%%%%%%%%%%%%%%%%%%%%%%%%%%%%%%%%%%%%%%%%%
\begin{titlepage}
$\mbox{ }$
\begin{flushright}
 \begin{tabular}{l}
  KEK-TH-1159\\
  June. 2007
 \end{tabular}
\end{flushright}
\vspace*{0cm}
\begin{Large}
 \vspace{2cm}
 \begin{center}
  {\bf Effective Actions of I\hspace{-.1em}IB Matrix Model
  on $S^3$}
\\
 \end{center}
\end{Large}
\vspace{1cm}
\begin{center}
 Hiromichi K{\sc aneko}$^{1)}$
 \footnote{E-mail address: kanekoh@post.kek.jp},
 Yoshihisa K{\sc itazawa}$^{1),2)}$
 \footnote{E-mail address: kitazawa@post.kek.jp}\\
 and
 Koichiro M{\sc atsumoto}$^{2)}$
 \footnote{E-mail address: kmatsumo@post.kek.jp}\\
\vspace{0.5cm}
  $^{1)}$ {\it High Energy Accelerator Research Organization (KEK),}\\
          {\it Tsukuba, Ibaraki 305-0801, Japan} \\
  $^{2)}$ {\it Department of Particle and Nuclear Physics,}\\
          {\it The Graduate University for Advanced Studies (SOKENDAI),}\\
          {\it Tsukuba, Ibaraki 305-0801, Japan}\\
\end{center}
\vspace{2cm}
\begin{abstract}
$S^3$ is a simple principle bundle which is locally $S^2\times S^1$.
It has been shown that such a space can be constructed
in terms of matrix models.
It has been also shown that such a space can be realized
by a generalized compactification procedure in the $S^1$ direction.
We investigate the effective action of supersymmetric  gauge theory on $S^3$
with an angular momentum cutoff and that of a matrix model compactification.
The both cases can be realized in a deformed I\hspace{-.1em}IB matrix model with a Myers term.
We find that the highly divergent contributions at the tree and 1-loop level
are sensitive to the UV cutoff. However the 2-loop level contributions
are universal since they are only logarithmically divergent.
We expect that the higher loop contributions are insensitive to the UV cutoff
since 3-dimensional gauge theory is super renormalizable.
\end{abstract}
\end{titlepage}
%
%\tableofcontents
%\listoffigures
%\listoftables
%
\newpage
%%%%%%%%%%%%%%%%%%%%%%%%%%%%%%%%%%%%%%%%%%%%%%%%%%%%%%%%%%%%%%%%%%%%%%%%%%%%%%%%
%%%%%%%%%%%%%%%%%%%%%%%%%%%%%%%%%%%% intro %%%%%%%%%%%%%%%%%%%%%%%%%%%%%%%%%%%%%
%%%%%%%%%%%%%%%%%%%%%%%%%%%%%%%%%%%%%%%%%%%%%%%%%%%%%%%%%%%%%%%%%%%%%%%%%%%%%%%%
\section{Introduction}
In our universe, there are many mysteries that remain to be understood
such as the selection mechanism of spacetime dimension,
gauge groups and matter contents.
It is important that we make progress to resolve these problems because we are increasing well
informed how our universe is formed.
Especially, we focus on the question why the 4-dimensionality of
spacetime is selected in our universe.

It is considered that superstring theory provides an effective tool
to explain the 4-dimensionality of spacetime. Since superstring
theory is a unified theory including the gravity, we may hope to
derive all physical predictions from the first principle.
Unfortunately, superstring theory suggests 10-dimensional
spacetime on the perturbative analysis. We believe that the
nonperturbative analysis of superstring theory is needed to explain
the 4-dimensionality of our universe. This question may be
addressed in the matrix models which are proposed for
nonperturbative formulations of superstring theory \cite{bfss,ikkt}.

I\hspace{-.1em}IB matrix model is a candidate for the
nonperturbative formulation of I\hspace{-.1em}IB superstring theory \cite{ikkt,aikktt}.
It is defined by the following action:
\begin{eqnarray}
 S_{{\rm I\hspace{-.1em}IB}}
  =-{\rm Tr}\left(\frac14\left[A_{\mu},A_{\nu}\right]^2
             +\frac12\bar{\psi}\Gamma^{\mu}\left[A_{\mu},\psi\right]\right),
\end{eqnarray}
where $A_{\mu}$ is a 10-dimensional vector and $\psi$ is a 10-dimensional
Majorana-Weyl spinor field respectively, and both fields are $N\times N$ Hermitian matrices.
There are considerable amount of investigations toward understanding the 4-dimensionality of spacetime by
using I\hspace{-.1em}IB matrix model.
For example, we may list the following studies: branched polymer picture \cite{aikkt},
complex phase effects \cite{nv,an} and mean-field approximations
\cite{ns,kkkms1,kkkms2}.
These studies seem to suggest that\\I\hspace{-.1em}IB matrix model
predicts 4-dimensionality of spacetime.

But it is difficult to analyze dynamics of I\hspace{-.1em}IB matrix
model in a generic spacetime. So we would like to understand
general mechanisms to single out the 4-dimensionality of
spacetime through the studies of concrete examples. We have
successfully constructed fuzzy homogeneous spaces using
I\hspace{-.1em}IB matrix model \cite{kitazawa}. We construct the
homogeneous spaces as $G/H$ where $G$ is a Lie group and $H$ is a
closed subgroup of $G$. When we give a background field to
$A_{\mu}$, we can examine the stability of this matrix
configurations by investigating the behavior of the effective action
under the change of some parameters of the background. We have
investigated the stabilities of fuzzy $S^2$ \cite{iktt-2003}, fuzzy
$S^2 \times S^2$ \cite{iktt-2004}, fuzzy $CP^2$ \cite{kkt-2006} and
fuzzy $S^2 \times S^2 \times S^2$ \cite{kkt-2005} in the past. We
have found that I\hspace{-.1em}IB matrix model favors the
configurations of 4-dimensionality and more symmetric manifolds.

In this paper, we investigate the 3-dimensional
sphere $S^3$ configuration.
We calculate the effective action of a deformed I\hspace{-.1em}IB
matrix model by introducing a Myers term up to the 2-loop level on $S^3$.
$S^3$ is a simple principle bundle which is locally $S^2\times S^1$.
It has been shown that such a space can be constructed
in terms of matrix models\cite{hkk}.
It has been also shown that such a space can be realized
by a generalized compactification procedure in the $S^1$ direction \cite{istt}.
We investigate the effective action of supersymmetric gauge theory on $S^3$
with an angular momentum cutoff and that of a matrix model compactification.
The both cases can be realized in a deformed I\hspace{-.1em}IB matrix model with a Myers term.
We find that the highly divergent contributions at the tree and 1-loop level
are sensitive to the UV cutoff.
However the 2-loop level contributions
are universal since they are only logarithmically divergent.
We expect that the higher loop contributions are insensitive to the UV cutoff
since 3-dimensional gauge theory is super renormalizable.

The organization of this paper is as follows:
In section 2, we review the properties of the $S^3$.
In section 3, we investigate the effective action of supersymmetric  gauge theory on $S^3$
with an angular momentum cutoff up to the 2-loop level.
In section 4, we investigate the corresponding effective action of a matrix model compactification.
In section 5,we conclude with discussions.
%
%\newpage
%%%%%%%%%%%%%%%%%%%%%%%%%%%%%%%%%%%%%%%%%%%%%%%%%%%%%%%%%%%%%%%%%%%%%%%%%%%%%%%%%%
%%%%%%%%%%%%%%%%%%%%%%%%%%%%%%%%%%%%%% sec.2 %%%%%%%%%%%%%%%%%%%%%%%%%%%%%%%%%%%%%
%%%%%%%%%%%%%%%%%%%%%%%%%%%%%%%%%%%%%%%%%%%%%%%%%%%%%%%%%%%%%%%%%%%%%%%%%%%%%%%%%%
\section{Derivatives on $S^3$}
In this section, we construct the derivatives on a 3-dimensional
sphere: $S^3$. First of all, we review basic properties of an $S^3$
\cite{biedenharn,br,cutkosky,sen}. $S^3$ is defined by the following
condition involving four Cartesian coordinates $x_n$:
\begin{eqnarray}
 \sum_{n=1}^4x_n^2=1.
 \label{cartesian_coordinates}
\end{eqnarray}
It is more convenient to introduce the following parametrization:
\begin{eqnarray}
 u=x_1+{\rm i}x_2=\cos\theta\,{\rm e}^{{\rm i}\phi}, \hspace{0.5cm}
 v=x_3+{\rm i}x_4=\sin\theta\,{\rm e}^{{\rm i}\tilde{\phi}},
 \label{s3-parameterization}
\end{eqnarray}
where $0\leq\theta\leq\pi/2$, $0\leq\phi<2\pi$ and
$0\leq\tilde{\phi}<2\pi$. The $S^3$ has an $SO(4)$ symmetry while
the $S^2$ has an $SO(3)$ symmetry. It is a well-known fact that the
$SO(4)$ is isomorphic to $SU(2) \times SU(2)$. Let $J_{1i}$ and
$J_{2i}$ denote the generators of each $SU(2)$ subgroup
respectively, where $i=1,2,3$. We define the generators of the
$SO(4)$ rotation group by the following linear combinations of the
$SU(2)$ generators:
\begin{eqnarray}
 &&M_i=J_{1i}+J_{2i}
      =-{\rm i}\epsilon_{ijk}x^j\frac{\partial}{\partial x_k}, \nonumber \\
 &&N_i=J_{1i}-J_{2i}
      ={\rm i}\left(x_4\frac{\partial}{\partial x^i}-x_i\frac{\partial}{\partial x_4}\right).
\end{eqnarray}
In fact, the operators $M_i$ represent the rotations around the
$x_i$ directions and the operators $N_i$ represent the rotations in
the $x_i$-$x_4$ planes. They obey the following commutation relations:
\begin{eqnarray}
 \left[M_i, M_j\right]={\rm i}\epsilon_{ijk} M^k, \hspace{0.5cm}
 \left[N_i, N_j\right]={\rm i}\epsilon_{ijk} M^k, \hspace{0.5cm}
 \left[M_i, N_j\right]={\rm i}\epsilon_{ijk} N^k.
\end{eqnarray}
$J_{1i}$ and $J_{2i}$ obey the following commutation relations:
\begin{eqnarray}
 &&\left[\mbox{\boldmath $J$}^2_1, J_{1i}\right]=0, \hspace{0.5cm}
   \left[J_{13}, J_{1\pm}\right]=\pm J_{1\pm}, \hspace{0.5cm}
   \left[J_{1+}, J_{1-}\right]=2J_{13},\nonumber \\
 &&\left[\mbox{\boldmath $J$}^2_2, J_{2i}\right]=0, \hspace{0.5cm}
   \left[J_{23}, J_{2\pm}\right]=\pm J_{2\pm}, \hspace{0.5cm}
   \left[J_{2+}, J_{2-}\right]=2J_{23},
\end{eqnarray}
where $J_{1\pm}=J_{11}\pm{\rm i}J_{12}$ and $J_{2\pm}=J_{21}\pm{\rm
i}J_{22}$. We can express these operators in terms of the
coordinates system (\ref{s3-parameterization}). For example,
\begin{eqnarray}
 J_{13}=-\frac{{\rm i}}{2}\left(\frac{\partial}{\partial\phi}
               +\frac{\partial}{\partial\tilde{\phi}}\right), \hspace{0.5cm}
 J_{23}=-\frac{{\rm i}}{2}\left(\frac{\partial}{\partial\phi}
               -\frac{\partial}{\partial\tilde{\phi}}\right).
\end{eqnarray}
The raising and lowering operators are
\begin{eqnarray}
 &&J_{1\pm}=\frac12{\rm e}^{\pm{\rm i}\phi}
            {\rm e}^{\pm{\rm i}\tilde{\phi}}
            \left(\pm \frac{\partial}{\partial\theta}
            +{\rm i}\tan\theta\frac{\partial}{\partial\phi}
            -{\rm i}\cot\theta\frac{\partial}{\partial\tilde{\phi}}\right),
              \nonumber \\
 &&J_{2\pm}=\frac12{\rm e}^{\pm{\rm i}\phi}
            {\rm e}^{\mp{\rm i}\tilde{\phi}}
            \left(\mp\frac{\partial}{\partial\theta}
            +{\rm i}\tan\theta\frac{\partial}{\partial\phi}
            +{\rm i}\cot\theta\frac{\partial}{\partial\tilde{\phi}}\right).
\end{eqnarray}
The $S^3$ is isomorphic to the $SU(2)$ group manifold as an element
of $SU(2)$ is represented by the Pauli matrices $\sigma_i$: $
U=x_4\mbox{\boldmath $1$}+{\rm i}\sum_{i=1}^3x_i\sigma_i$. When we
impose the condition of the unitarity:
$U^{\dagger}U=UU^{\dagger}=\mbox{\boldmath $1$}$ and speciality:
${\det}U=1$, we obtain the equation (\ref{cartesian_coordinates}).

This space is a homogeneous space as we recall the following relation:
\begin{eqnarray}
 SO(4)/SU(2)=SU(2)\sim S^3.
\end{eqnarray}
The homogeneous space is represented as $G/H$, where $G$ is a Lie
group and $H$ is a closed subgroup of $G$. We can construct the
$S^3$ by using the subgroup $SU(2)$ of the $SO(4)$. Let us choose
the basis vectors as follows:
\begin{eqnarray}
 \mbox{\boldmath $\hat{e}$}^{\,(1)}_1&=&(x_4,x_3,-x_2,-x_1), \nonumber \\
 \mbox{\boldmath $\hat{e}$}^{\,(1)}_2&=&(-x_3,x_4,x_1,-x_2), \\
 \mbox{\boldmath $\hat{e}$}^{\,(1)}_3&=&(x_2,-x_1,x_4,-x_3), \nonumber
\end{eqnarray}
where $(a,b,c,d)$ denotes the Cartesian components of a vector. We
find that the components of a vector $\mbox{\boldmath $V$}$ is
denoted by $V^{\,(1)}_i=\mbox{\boldmath $V$}\cdot\mbox{\boldmath
$e$}^{\,(1)}_i$, and then the derivatives $\partial^{\,(1)}_i$ along
these axes are denoted by
\begin{eqnarray}
 \partial^{\,(1)}_i=-2{\rm i}J_{1i}, \hspace{0.5cm}
 \left[\partial^{\,(1)}_i,\partial^{\,(1)}_j\right]=2\epsilon_{ijk}\partial^{\,(1)k}.
\end{eqnarray}
The derivatives on the $S^3$ is constructed by a Lie algebra of the $SU(2)$.
On the other hand, we can choose another different basis vectors as follows:
\begin{eqnarray}
 \mbox{\boldmath $\hat{e}$}^{\,(2)}_1&=&(x_4,-x_3,x_2,-x_1), \nonumber \\
 \mbox{\boldmath $\hat{e}$}^{\,(2)}_2&=&(x_3,x_4,-x_1,-x_2), \\
 \mbox{\boldmath $\hat{e}$}^{\,(2)}_3&=&(-x_2,x_1,x_4,-x_3). \nonumber
\end{eqnarray}
In the same way, we can construct another set of derivatives on
$S^3$
\begin{eqnarray}
 \partial^{\,(2)}_i=2{\rm i}J_{2i}, \hspace{0.5cm}
 \left[\partial^{\,(2)}_i,\partial^{\,(2)}_j\right]=-2\epsilon_{ijk}\partial^{\,(2)k}.
\end{eqnarray}
In the parameterization (\ref{s3-parameterization}) of the $S^3$, we find
the Laplacian on the $S^3$ is as follows:
\begin{eqnarray}
 \bigtriangleup_3
  =\frac{1}{\sin\theta\cos\theta}
   \frac{\partial}{\partial\theta}
   \left(\sin\theta\cos\theta\frac{\partial}{\partial\theta}\right)
   +\frac{1}{\cos^2\theta}
   \frac{\partial^2}{\partial\phi^2}
   +\frac{1}{\sin^2\theta}
   \frac{\partial^2}{\partial\tilde{\phi}^2}.
\end{eqnarray}
It is easy to recognize that the Casimir operators act
as the Laplacian on the $S^3$:
\begin{eqnarray}
 \mbox{\boldmath $J$}^2_1
  =\mbox{\boldmath $J$}^2_2
  =-\frac14\bigtriangleup_3.
\end{eqnarray}

We may consider an eigenvalue equation of the Laplacian on the $S^3$
as follows:
\begin{eqnarray}
 \bigtriangleup_3Y(\theta,\phi,\tilde{\phi})
  =-\lambda Y(\theta,\phi,\tilde{\phi}),
 \label{eigenvalue_eq}
\end{eqnarray}
where $Y(\theta,\phi,\tilde{\phi})$ represents the solutions of this
equation. We can solve this equation by separating the variables
completely: $Y(\theta,\phi,\tilde{\phi})=\Theta(\theta)\exp({\rm
i}m\phi+{\rm i}\tilde{m}\tilde{\phi})$. We find the following
differential equation:
\begin{eqnarray}
 \left(1-z^2\right)\frac{d^2P(z)}{dz^2}
  -2z\frac{dP(z)}{dz}
  +\left(\frac14\lambda
    -\frac{m^2+\tilde{m}^2}{2}\frac{1}{1-z^2}
    +\frac{m^2-\tilde{m}^2}{2}\frac{z}{1-z^2}\right)P(z)=0,
  \nonumber \\
 \label{diff_eq}
\end{eqnarray}
where $z=\cos2\theta$ and $P(z)=\Theta(\theta)$. The $m$ and
$\tilde{m}$ take integers because $\exp({\rm i}m\phi+{\rm
i}\tilde{m}\tilde{\phi})$ has a periodicity of $2\pi$. If we
consider the case $m=\tilde{m}\equiv k$, we obtain the Legendre's
differential equation:
\begin{eqnarray}
 \left(1-z^2\right)\frac{d^2P(z)}{dz^2}
  -2z\frac{dP(z)}{dz}
  +\left(\frac14\lambda-\frac{k^2}{1-z^2}\right)P(z)=0.
\end{eqnarray}
We can solve this equation by a series expansion. We obtain Legendre
polynomials as the solutions with the eigenvalues: $\lambda=n(n+2)$
where $n$ is a positive integer. We can subsequently operate the
raising and lowering operators $J_{1\pm}$ and $J_{2\pm}$ to the
Legendre polynomials and obtain the complete solutions of
(\ref{diff_eq}) for $m\neq\tilde{m}$. In this way we find that the
solutions of the differential equation (\ref{diff_eq}) are the
spherical harmonics on the $S^3$:
\begin{eqnarray}
 Y^{\;n}_{m\tilde{m}}(\theta,\phi,\tilde{\phi})
  =\sqrt{n+1}\,{\rm e}^{{\rm i}m\phi+{\rm i}\tilde{m}\tilde{\phi}}\,
  d^{\,\frac{n}{2}}_{\frac12(m+\tilde{m}),\frac12(m-\tilde{m})}\left(2\theta\right),
\end{eqnarray}
where the $d$ functions are given in terms of the Jacobi polynomials
\cite{edmonds,vmk}, $-n/2\leq(m+\tilde{m})/2\leq n/2$ and
$-n/2\leq(m-\tilde{m})/2\leq n/2$. When the $J_{13}$ and $J_{23}$
operate on the spherical harmonics $Y^{\; n}_{m\tilde{m}}(\Omega)$
on the $S^3$, we obtain eigenvalues $\left(m+\tilde{m}\right)/2$ and
$\left(m-\tilde{m}\right)/2$ respectively. They satisfy the
following orthogonality condition:
\begin{eqnarray}
 \int\!d\Omega\,
  {Y^{\;n_1}_{m_1\tilde{m}_1}}^{\ast}\left(\Omega\right)
  Y^{\;n_2}_{m_2\tilde{m}_2}\left(\Omega\right)
  =\delta_{n_1n_2}\delta_{m_1m_2}\delta_{\tilde{m}_1\tilde{m}_2}.
\end{eqnarray}
We define the volume element $d\Omega$ as follows:
\begin{eqnarray}
 d\Omega=\frac{1}{2\pi^2}\sin\theta\cos\theta\,d\theta\,d\phi\,d\tilde{\phi}.
\end{eqnarray}
%
%\newpage
%%%%%%%%%%%%%%%%%%%%%%%%%%%%%%%%%%%%%%%%%%%%%%%%%%%%%%%%%%%%%%%%%%%%%%%%%%%%%%%%%
%%%%%%%%%%%%%%%%%%%%%%%%%%%%%%%%%%%%% sec.3 %%%%%%%%%%%%%%%%%%%%%%%%%%%%%%%%%%%%%
%%%%%%%%%%%%%%%%%%%%%%%%%%%%%%%%%%%%%%%%%%%%%%%%%%%%%%%%%%%%%%%%%%%%%%%%%%%%%%%%%
\section{Effective action on $S^3$}
In this section, we investigate the effective action of a deformed
I\hspace{-.1em}IB matrix model on the $S^3$ background. In order to
obtain $S^3$ as classical solutions of I\hspace{-.1em}IB matrix
model, we deform the action of I\hspace{-.1em}IB matrix model by
adding a Myers term as follows:
\begin{eqnarray}
 S=-{\rm Tr}\left(\frac14\left[A_{\mu},A_{\nu}\right]^2
       +\frac12\bar{\psi}\Gamma^{\mu}\left[A_{\mu},\psi\right]
       -\frac{{\rm i}}{3}f_{\mu\nu\rho}\left[A^{\mu},A^{\nu}\right]A^{\rho}\right),
 \label{deformed_IIB}
\end{eqnarray}
where $f_{\mu\nu\rho}$ is the structure constant of $SU(2)$. When we
assume $\psi=0$, we obtain the equation of motion for $A_{\mu}$:
\begin{eqnarray}
 \Bigl[A_{\mu}, [A^{\mu},A_{\nu}]\Bigr]
  -{\rm i}f_{\mu\nu\rho}[A^{\mu},A^{\rho}]=0.
\end{eqnarray}
The nontrivial classical solutions of this equation of motion are
\begin{eqnarray}
 A_a=t_a,\hspace{0.5cm}{\rm other}\hspace{0.3cm}A_{\mu}=0,
\end{eqnarray}
where $t_a$'s satisfy the Lie algebra. To calculate the effective
action of the deformed I\hspace{-.1em}IB matrix model, we decompose
the matrices $A_{\mu}$ and $\psi$ into the backgrounds and the
quantum fluctuations as follows:
\begin{eqnarray}
 A_{\mu}&=&p_{\mu}+a_{\mu}, \nonumber \\
 \psi&=&\chi+\varphi,
\end{eqnarray}
where $p_{\mu}$ and $\chi$ are the backgrounds, and $a_{\mu}$ and $\varphi$
are the quantum fluctuations.
When we expand the action (\ref{deformed_IIB}) around the quantum
fluctuations, we also add the gauge fixing term and the Faddeev-Popov ghost
term as follows:
\begin{eqnarray}
 S_{\rm g.f.}&=&-\frac12{\rm Tr}\left([p_{\mu},a^{\mu}]^2\right), \nonumber \\
 S_{\rm F.P.}&=&{\rm Tr}\left(b\Bigl[p_{\mu},[p^{\mu}+a^{\mu},c]\Bigr]\right),
\end{eqnarray}
where $c$ and $b$ are ghosts and anti-ghosts, respectively.
In this way we find that the following action:
\begin{eqnarray}
 \tilde{S}
   &=&S+S_{\rm g.f.}+S_{\rm F.P.} \nonumber \\
   &=&-{\rm Tr}\left(\frac14[p_{\mu},p_{\nu}]^2
             +\frac12\bar{\chi}\Gamma^{\mu}[p_{\mu},\chi]
             -\frac{{\rm i}}{3}f_{\mu\nu\rho}\left[p^{\mu},p^{\nu}\right]p^{\rho}\right.
                \nonumber \\
   &&-\,a_{\nu}\!\left(\Bigl[p_{\mu},[p^{\mu},p^{\nu}]\Bigr]
             +\frac12\{\bar{\chi}\Gamma^{\nu},\chi\}\right)
             +\bar{\chi}\Gamma^{\mu}[p_{\mu},\varphi]
             -{\rm i}f_{\mu\nu\rho}\left[p^{\mu},p^{\nu}\right]a^{\rho}
                 \nonumber \\
   &&+\,\frac12[p_{\mu},a_{\nu}]^2
     +[p_{\mu},p_{\nu}][a^{\mu},a^{\nu}]
     -b\Bigl[p_{\mu},[p^{\mu},c]\Bigr]
     +\frac12\bar{\varphi}\Gamma^{\mu}[p_{\mu},\varphi]
     +\bar{\chi}\Gamma^{\mu}[a_{\mu},\varphi]
     +{\rm i}f_{\mu\nu\rho}a^{\mu}\left[p^{\rho},a^{\nu}\right]
           \nonumber \\
   &&\left.+\,[p_{\mu},a_{\nu}][a^{\mu},a^{\nu}]
      -b\Bigl[p_{\mu},[a^{\mu},c]\Bigr]
      +\frac12\bar{\varphi}\Gamma^{\mu}[a_{\mu},\varphi]
      -\frac{{\rm i}}{3}f_{\mu\nu\rho}\left[a^{\mu},a^{\nu}\right]a^{\rho}
      +\frac14[a_{\mu},a_{\nu}]^2\right). \nonumber \\
\end{eqnarray}

Since we investigate the effective action on the $S^3$ background,
we may substitute the derivatives ${\rm i}\,\partial^{\,(1)}_i$ on
the $S^3$ for the backgrounds as follows:
\begin{eqnarray}
 &&p_i={\rm i}\,\partial^{\,(1)}_i=\alpha J_{1i}, \hspace{0.5cm}
   {\rm other}\hspace{0.3cm}p_{\mu}=0, \nonumber \\
 &&\chi=0,
 \label{bg-derivative}
\end{eqnarray}
where $\alpha$ is a scale factor and $i=1,2,3$. In \cite{hkk}, the
authors have found that the bosonic part $A_{\mu}$ of
I\hspace{-.1em}IB matrix model can be interpreted as differential
operators on the principle bundles. One of the goals of our
investigations is to obtain deeper understanding of such an
interpretation through a concrete example: $S^3$.

When the backgrounds is a flat space, we expand the quantum
fluctuations by a plane wave. Since we consider the $S^3$
background, it is natural to expand the quantum fluctuations by a
spherical harmonics on the $S^3$:
\begin{eqnarray}
 a_{\mu}&=&\sum_{nm\tilde{m}}a^{\;\;\;n}_{\mu m\tilde{m}}
                 Y^{\;n}_{m\tilde{m}}\left(\Omega\right), \nonumber \\
 \varphi&=&\sum_{nm\tilde{m}}\varphi^{\;\;n}_{m\tilde{m}}
                 Y^{\;n}_{m\tilde{m}}\left(\Omega\right),
 \label{qf-derivative}
\end{eqnarray}
where $a^{\;\;\;n}_{\mu m\tilde{m}}$ and
$\varphi^{\;\;n}_{m\tilde{m}}$ are expansion coefficients. In the
same way, ghosts and anti-ghosts fields are expanded by the
spherical harmonics on the $S^3$:
\begin{eqnarray}
 c&=&\sum_{nm\tilde{m}}c^{\;\;n}_{m\tilde{m}}\,
    Y^{\;n}_{m\tilde{m}}\left(\Omega\right), \nonumber \\
 b&=&\sum_{nm\tilde{m}}b^{\;\;n}_{m\tilde{m}}\,
    Y^{\;n}_{m\tilde{m}}\left(\Omega\right),
    \label{gh-derivative}
\end{eqnarray}
The structure constants $f_{\mu\nu\rho}$ are given as follows:
\begin{eqnarray}
 f_{ijk}=\alpha\epsilon_{ijk}, \hspace{0.5cm}
  {\rm other} \hspace{0.3cm} f_{\mu\nu\rho}=0.
 \label{sc-derivative}
\end{eqnarray}
The gauge fixed action around the backgrounds (\ref{bg-derivative})
take the following form:
\begin{eqnarray}
 \tilde{S}
  &=&-{\rm Tr}\left(\frac14\left[p_i,p_j\right]^2
     -\frac{{\rm i}}{3}f_{ijk}[p^i,p^j]p^k\right. \nonumber \\
  &&+\,\frac12\left[p_i,a_{\mu}\right]^2
    -b\Bigl[p_i,[p^i,c]\Bigr]
    +\frac12\bar{\varphi}\Gamma^i\left[p_i,\varphi\right]
     \nonumber \\
  &&\left.+\,[p_i,a_{\mu}][a^i,a^{\mu}]
     -b\Bigl[p_i,[a^i,c]\Bigr]
     +\frac12\bar{\varphi}\Gamma^{\mu}\left[a_{\mu},\varphi\right]
     -\frac{{\rm i}}{3}f_{ijk}[a^i,a^j]a^k
     +\frac14\left[a_{\mu},a_{\nu}\right]^2\right). \nonumber \\
 \label{tilde-s}
\end{eqnarray}
Using the above action $\tilde{S}$, we can evaluate the effective
action $W$ of the deformed I\hspace{-.1em}IB matrix model on the
$S^3$ background as follows in a background gauge method:
\begin{eqnarray}
 W=-\log\!\int\!da\,d\varphi\,dc\,db\;{\rm e}^{-\tilde{S}}.
\end{eqnarray}

Firstly, we evaluate the effective action at the tree level
\begin{eqnarray}
 W_{\rm tree}
  &=&-{\rm Tr}\left(\frac14\left[p_i,p_j\right]^2
               -\frac{{\rm i}}{3}f_{ijk}[p^i,p^j]p^k\right) \nonumber \\
  &=&-\frac{\alpha^4}{6}{\rm Tr}\left(J_{1i}\right)^2.
\end{eqnarray}
In the last step, we substitute the derivatives: ${\rm
i}\,\partial^{\;(1)}_i=\alpha J_{1i}$ on the $S^3$ for the
backgrounds $p_i$ of the deformed I\hspace{-.1em}IB matrix model. We
evaluate the effective action of the deformed I\hspace{-.1em}IB
matrix model by taking the continuous limit as follows:
\begin{eqnarray}
 {\rm Tr}X
  \longrightarrow
  \int\!d\Omega\left<\Omega|\,X\,|\Omega\right>.
\end{eqnarray}
We may consider that this limit corresponds to a semi-classical
limit. We evaluate the trace of the Casimir operator as:
\begin{eqnarray}
 {\rm Tr}\left(J_{1i}\right)^2
  &\longrightarrow&
   -\frac14\int\!d\Omega\left<\Omega|\bigtriangleup_3|\Omega\right>
         \nonumber \\
  &=&\frac14\int\!d\Omega\sum_{nm\tilde{m}}
    n(n+2)\left<\Omega|n,m,\tilde{m}\right>\!\!\left<n,m,\tilde{m}|\Omega\right>
          \nonumber \\
  &=&\frac14\sum_nn(n+2)(n+1)^2.
\end{eqnarray}
In the second line, we have used the complete set of the eigenstates
for the Laplacian on the $S^3$:
$\sum_{nm\tilde{m}}\left|n,m,\tilde{m}\right>\!\!\left<n,m,\tilde{m}\right|=\mbox{\boldmath$1$}$.
In the last line, we have used the fact that
$Y^{\;n}_{m\tilde{m}}\left(\Omega\right)=\left<\Omega|n,m,\tilde{m}\right>$
and the degeneracy factors coming from $m$ and $\tilde{m}$ are
$(n+1)^2$. Therefore,
\begin{eqnarray}
 W_{\rm tree}\longrightarrow-\frac{\alpha^4}{24}\sum_nn(n+2)(n+1)^2.
\end{eqnarray}
Since $n$ takes a positive integer, the summation over $n$ is
formally divergent. It is necessary to impose a cutoff at a some
large but finite $n$. When impose a cutoff at $n=l$, the tree level
effective action is evaluated as follows:
\begin{eqnarray}
 W_{\rm tree}
  \longrightarrow
  -\frac{\alpha^4}{24}
  \sum_{n=1}^{l}
  n(n+2)(n+1)^2
  \sim
  -\frac{\alpha^4}{24}\,
  l^5.
  \label{w-tree}
\end{eqnarray}

Secondly, we evaluate the effective action at the 1-loop level as
follows:
\begin{eqnarray}
 W_{\rm 1-loop}
  &=&\frac12{\rm Tr}\log\left(P_i^2\,\delta^{\mu\nu}\right)
     -{\rm Tr}\log\left(P_i^2\right) \nonumber \\
  &&-\frac14{\rm Tr}\log\Biggl[\left(P_i^2+\frac{{\rm i}}{2}F_{ij}\Gamma^{ij}\right)\!\!
       \left(\frac{1+\Gamma_{11}}{2}\right)\Biggr],
 \label{w_1-loop}
\end{eqnarray}
where
\begin{eqnarray}
 &&\left[p_i, X\right]=P_iX, \nonumber \\
 &&\left[f_{ij},X\right]=F_{ij}X,\;\;\;{\rm i}f_{ij}=\left[p_i,p_j\right].
\end{eqnarray}
The first and the second terms are the bosonic contributions while
the third term is the fermionic contribution. We have to include a
projection operator $(1+\Gamma_{11})/2$ in the fermionic part
because we consider a 10-dimensional Majorana-Weyl spinor field. We
expand the third term of the equation (\ref{w_1-loop}) into the
power series of $P_i$ and $F_{ij}$. In this way, we obtain the
leading term of the 1-loop level effective action as follows:
\begin{eqnarray}
 W_{\rm 1-loop}
  \sim-{\rm Tr}\left(\frac{1}{P_i^2}\right)^2\!\!F_{ij}F^{ji}
  ={\rm Tr}\frac{2\alpha^2}{P_i^2}.
\end{eqnarray}
When we impose the same cutoff procedure: $n\leq l$, we obtain the
effective action at the 1-loop level as follows:
\begin{eqnarray}
 W_{\rm 1-loop}\longrightarrow 8\sum_{n=1}^{l}\frac{(n+1)^2}{n(n+2)}
  \sim 8\,l.
\end{eqnarray}

Finally, we evaluate the effective action at the 2-loop level due to
planar diagrams.
We describe the detailed calculations of the 2-loop effective action in
appendix B.
The effective action at the 2-loop level is:
\begin{eqnarray}
 W_{\rm 2-loop}
 =\frac{2304}{\alpha^4}\sum_{n_1n_2n_3}
  \frac{(n_1+1)(n_2+1)(n_3+1)}{n_1(n_1+2)n_2(n_2+2)n_3(n_3+2)}.
\end{eqnarray}
We impose the same the cutoff procedure of the summations over
$n_1$, $n_2$ and $n_3$:
\begin{eqnarray}
 \sum_{n_1=1}^l\sum_{n_2=1}^l\sum_{n_3=1}^l
 \frac{(n_1+1)(n_2+1)(n_3+1)}{n_1(n_1+2)n_2(n_2+2)n_3(n_3+2)}
 \equiv f\left(l\right).
\end{eqnarray}
where we recall the following selection rules that
$\left|n_1-n_2\right|\leq n_3\leq n_1+n_2$ and $n_1+n_2+n_3$ must be
even numbers.
To the leading logarithmic order, we can analytically evaluate it:
\begin{eqnarray}
 f\left(l\right)\sim\frac{\pi^2}{4}\log l,
  \label{sum-n1_n2_n3}
\end{eqnarray}
We illustrate the comparison between the numerical evaluation and the analytic
expression (\ref{sum-n1_n2_n3})  in Fig. 1.
We find that the analytic expression is valid to the leading logarithmic order
as the slope of the two lines are identical.
\begin{figure}[htbp]
 \begin{center}
  \includegraphics[scale=1.0]{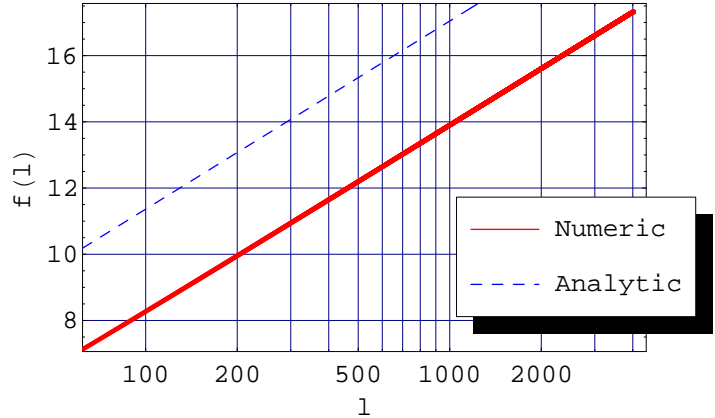}
  \caption{The numerical and analytic calculations of the 2-loop corrections.}
  \label{fig:numeric-analytic}
 \end{center}
\end{figure}
We thus conclude that the 2-loop level effective action is
\begin{eqnarray}
 W_{\rm 2-loop}\sim\frac{576\pi^2}{\alpha^4}\log l.
\end{eqnarray}
In this way we can summarize the effective action with an angular
momentum cutoff $l$ up to the 2-loop level:
\begin{eqnarray}
 W\sim
  -\frac{\alpha^4}{24}\,l^5
  +8\,l
  +\frac{576\pi^2}{\alpha^4}\log l.
  \label{eff_action-derivative}
\end{eqnarray}
While the tree and 1-loop contributions are highly divergent, the
2-loop contribution is only logarithmically divergent.This result
is consistent with the fact that 3-dimensional gauge theory is super
renormalizable and we expect that the higher loop contributions are
finite. Since $1/\alpha^4$ acts as the loop expansion parameter, we
need to assume $\alpha\sim{\cal O}\left(1\right)$. We then conclude
that the effective action of the deformed I\hspace{-.1em}IB matrix
model on the $S^3$ is stable against the quantum corrections as it
is dominated by the tree level contribution.
%
%\newpage
%%%%%%%%%%%%%%%%%%%%%%%%%%%%%%%%%%%%%%%%%%%%%%%%%%%%%%%%%%%%%%%%%%%%%%%%%%%%%%%%%
%%%%%%%%%%%%%%%%%%%%%%%%%%%%%%%%%%%%%%% sec.4 %%%%%%%%%%%%%%%%%%%%%%%%%%%%%%%%%%%
%%%%%%%%%%%%%%%%%%%%%%%%%%%%%%%%%%%%%%%%%%%%%%%%%%%%%%%%%%%%%%%%%%%%%%%%%%%%%%%%%
\section{I\hspace{-.1em}IB matrix model compactification on $S^3$}
There is an interesting construction of an $S^3$ background in matrix models recently.
In \cite{istt}, the authors have concluded that the $S^3$ is realized by
three matrices.
They have proved the following two relations between the vacua
of different gauge theories:
\begin{description}
 \item[(i)] The theory around each vacuum of  super Yang-Mills
            theory on $S^2$ (${\rm SYM}_{S^2}$) is
            equivalent to the theory around a certain vacuum of a
            matrix model.
 \item[(i\hspace{-.1em}i)]  The theory around each vacuum of the
            super Yang-Mills theory on $S^3/Z_k$ (${\rm
            SYM}_{S^3/Z_k}$) is equivalent to the theory around
            a certain vacuum of ${\rm SYM}_{S^2}$ with
            periodic identifications.
\end{description}
They selected the following nontrivial vacua of ${\rm SYM}_{S^2}$:
\begin{align}
 \Phi=\frac{\mu}{2}
  \begin{pmatrix}
   \rotatebox[origin=tl]{-35}
   {$
   \overbrace{\rotatebox[origin=c]{35}{$\alpha_1$} \;\cdots\;
   \rotatebox[origin=c]{35}{$\alpha_1$}}^{\rotatebox{35}{$N_1$}} \;\;\;
   \overbrace{\rotatebox[origin=c]{35}{$\alpha_2$} \;\cdots\;
   \rotatebox[origin=c]{35}{$\alpha_2$}}^{\rotatebox{35}{$N_2$}}
   \;\;\;\cdots\;\;\;
   \overbrace{\rotatebox[origin=c]{35}{$\alpha_T$} \;\cdots\;
   \rotatebox[origin=c]{35}{$\alpha_T$}}^{\rotatebox{35}{$N_T$}}
   $}
  \end{pmatrix},
 \label{vacuum-s2}
\end{align}
\begin{eqnarray}
 &&\hspace{-0.9cm}A_{\theta}=0, \nonumber\\
 &&\hspace{-0.9cm}A_{\phi}=\left\{
             \begin{array}{ll}
              \frac{1}{\mu}\left(1-\cos\theta\right)\Phi & \hspace{0.5cm}
               {\rm for}\,\,\,\,0\leq\theta<\frac{\pi}{2}+\epsilon \\
              -\frac{1}{\mu}\left(1+\cos\theta\right)\Phi & \hspace{0.5cm}
               {\rm for}\,\,\,\,\frac{\pi}{2}-\epsilon<\theta\leq\pi \\
             \end{array}
            \right.,
\end{eqnarray}
where $\alpha_s$'s ($s=1,\cdots,T$) parameterize monopole charges:
\begin{eqnarray}
 q_{st}=\frac12\left(\alpha_s-\alpha_t\right),
\end{eqnarray}
and all $\alpha_s$'s are different. Additionally we have the following relation
\begin{eqnarray}
 N_1+\cdots+N_T=\tilde{N}.
\end{eqnarray}
The radius of the $S^2$ is fixed to be $1/\mu$.
The fields in ${\rm SYM}_{S^2}$ are split into the blocks of
$N_s\times N_t$ rectangular matrices around the above nontrivial vacua.

On the other hand, a vacuum of a matrix model is represented as follows:
\begin{eqnarray}
 Y_i=-\mu L_i,
\end{eqnarray}
where
\begin{align}
 L_i=
  \begin{pmatrix}
   \rotatebox[origin=tl]{-35}
   {$
   \overbrace{\rotatebox[origin=c]{35}{$L_{i}^{[j_{1}]}$} \;\cdots\;
   \rotatebox[origin=c]{35}{$L_{i}^{[j_{1}]}$}}^{\rotatebox{35}{$N_1$}} \;\;\;
   \overbrace{\rotatebox[origin=c]{35}{$L_i^{[j_2]}$} \;\cdots\;
   \rotatebox[origin=c]{35}{$L_i^{[j_2]}$}}^{\rotatebox{35}{$N_2$}}
   \;\;\;\cdots\;\;\;
   \overbrace{\rotatebox[origin=c]{35}{$L_{i}^{[j_{T}]}$} \;\cdots\;
   \rotatebox[origin=c]{35}{$L_{i}^{[j_{T}]}$}}^{\rotatebox{35}{$N_T$}}
   $}
  \end{pmatrix}.
 \label{vacuum-pwmm}
\end{align}
The $L_i$ is a reducible $\hat{N}$-dimensional
representation of $SU(2)$, and obeys the following commutation
relation:
\begin{eqnarray}
 \left[L_i,L_j\right]={\rm i}\epsilon_{ijk}L^k,
\end{eqnarray}
where
\begin{eqnarray}
 (2j_1+1)N_1+(2j_2+1)N_2+\cdots+(2j_T+1)N_T=\hat{N}.
\end{eqnarray}
$L_i^{[j_s]}$ ($s=1,\cdots,T$) is the $(2j_s+1)\times
(2j_s+1)$ spin $j_s$ representation of $SU(2)$, and obeys the commutation relation:
\begin{eqnarray}
 \left[L_i^{[j_s]},L_j^{[j_s]}\right]={\rm i}\epsilon_{ijk}L^{[j_s]k}.
\end{eqnarray}
The Casimir operator of $L_i^{[j_s]}$ is that
\begin{eqnarray}
 L_i^{[j_s]}L^{[j_s]i}=j_s\left(j_s+1\right)\mbox{\boldmath $1$}_{2j_s+1}.
\end{eqnarray}
This vacuum (\ref{vacuum-pwmm}) can be interpreted as a set of
coincident $N_s$ fuzzy spheres with the radii $\mu\sqrt{j_s(j_s+1)}$,
where all the fuzzy spheres are concentric.

In order to prove the above two relations, they expand the theories around
various vacua using appropriate spherical harmonics respectively.
The spherical harmonics $Y_{JM\tilde{M}}$ on an $S^3$ is relevant for ${\rm SYM}_{S^3/Z_k}$
($J=n/2$, $M=(m+\tilde{m})/2$ and $\tilde{M}=(m-\tilde{m})/2$ in our convention)
where $J=0,1/2,1,\cdots$, $M=-J,-J+1,\cdots,J-1,J$ and
$\tilde{M}=-J,-J+1,\cdots,J-1,J$.
The monopole harmonic function $\tilde{Y}_{JMq_{st}}$ is used to expand
around the background of ${\rm SYM}_{S^2}$, where
$J=\left|q_{st}\right|,\left|q_{st}\right|+1,\left|q_{st}\right|+2,\cdots$,
and $M=-J,-J+1,\cdots,J-1,J$.
For a matrix model, fuzzy sphere harmonics $\hat{Y}_{JM}^{(j_sj_t)}$:
the harmonic function on a set of fuzzy spheres with different radii, is used
where $J=\left|j_s-j_t\right|,\left|j_s-j_t\right|+1,\cdots,j_s+j_t$ and
$M=-J,-J+1,\cdots,J-1,J$.
The equivalence (i) is proved when the following conditions are imposed
on the parameters of the vacua of ${\rm SYM}_{S^2}$ and the vacua of
a matrix model:
\begin{eqnarray}
 &&j_s-j_t=\frac12\left(\alpha_s-\alpha_t\right)=q_{st}, \nonumber \\
 &&j_s,\,j_t\longrightarrow\infty.
\end{eqnarray}

The equivalence (i\hspace{-.1em}i) is proved under the following conditions
\begin{eqnarray}
 &&\alpha_s=sk, \hspace{0.5cm} N_s=N, \nonumber \\
 &&s=1,\cdots,\infty.
\end{eqnarray}
Additionally, they identify the fuzzy spheres by imposing the
periodicity on the $(s,t)$ blocks and by factoring out the overall
factor. Combining the equivalences (i) and (i\hspace{-.1em}i), they
have concluded that the theory around the trivial vacua of ${\rm
SYM}_{S^3/Z_k}$ is equivalent to the theory around the vacua of a
matrix model. In order to draw the above conclusion, the following
conditions are necessary:
\begin{eqnarray}
 &&j_s-j_t=\frac{k}{2}\left(s-t\right)=q_{st}, \hspace{0.5cm} N_s=N, \nonumber \\
 &&j_s,\,j_t\longrightarrow\infty, \hspace{0.5cm} s,\,t=1,\cdots,\infty.
 \label{s3-PWMM}
\end{eqnarray}
The condition that $j_s-j_t=k(s-t)/2$ can be also written as
$2j_s+1=N_0+ks$, where $N_0$ is a positive integer.
The condition that $j_s\rightarrow\infty$ corresponds to
$N_0\rightarrow\infty$.

We make use of the work \cite{istt} to make a connection between super
Yang-Mills theory on an $S^3$ background and I\hspace{-.1em}IB matrix model on an
$S^3$ background.
We expand the deformed I\hspace{-.1em}IB\\matrix model action
(\ref{deformed_IIB}) around the backgrounds in an analogous way in section 3.
We introduce the matrix $Y_i$ as the backgrounds of the deformed I\hspace{-.1em}IB
matrix model:
\begin{eqnarray}
 &&p_i=Y_i=-\mu L_i=\beta L_i,\hspace{0.5cm}{\rm other}\hspace{0.3cm}p_{\mu}=0,
  \nonumber \\
 &&\chi=0.
  \label{bg-matrix}
\end{eqnarray}
where $\beta$ is a scale factor and $i=1,2,3$.
We make a mode expansion of the quantum fluctuations using the fuzzy
sphere harmonics:
\begin{eqnarray}
 &&a_{\mu}^{(s,t)}=\sum_{J=|j_s-j_t|}^{j_s+j_t}
   \sum_{M=-J}^{J}a_{\mu JM}^{(s,t)}\otimes\hat{Y}_{JM}^{(j_sj_t)},
   \nonumber \\
 &&\varphi^{(s,t)}=\sum_{J=|j_s-j_t|}^{j_s+j_t}
   \sum_{M=-J}^{J}\varphi^{(s,t)}_{JM}\otimes\hat{Y}_{JM}^{(j_sj_t)},
   \label{qf-matrix}
\end{eqnarray}
where the suffix $(s,t)$ represents the $(s,t)$ block in an
$\hat{N}\times\hat{N}$ matrix and $s,t=1,\cdots,T$. The coefficients
of the mode expansion $a_{\mu JM}^{(s,t)}$ and
$\varphi^{(s,t)}_{JM}$ are $N_s\times N_t$ matrices and the fuzzy
sphere harmonics $\hat{Y}_{JM}^{(j_sj_t)}$ is a
$\left(2j_s+1\right)\times\left(2j_t+1\right)$ matrix. Therefore,
$a_{\mu}^{(s,t)}$ and $\varphi^{(s,t)}$ are
$N_s\left(2j_s+1\right)\times N_t\left(2j_t+1\right)$ rectangular
matrices. Similarly, ghosts and anti-ghosts fields are expanded by
the fuzzy sphere harmonics:
\begin{eqnarray}
 &&c^{(s,t)}=\sum_{J=|j_s-j_t|}^{j_s+j_t}
   \sum_{M=-J}^{J}c_{JM}^{(s,t)}\otimes\hat{Y}_{JM}^{(j_sj_t)}, \nonumber \\
 &&b^{(s,t)}=\sum_{J=|j_s-j_t|}^{j_s+j_t}
   \sum_{M=-J}^{J}b_{JM}^{(s,t)}\otimes\hat{Y}_{JM}^{(j_sj_t)},
   \label{gh-matrix}
\end{eqnarray}
where $c_{JM}^{(s,t)}$ and $b_{JM}^{(s,t)}$ are the expansion
coefficients. As for structure constant, we adopt that
\begin{eqnarray}
 f_{ijk}=\beta\epsilon_{ijk},\hspace{0.5cm}{\rm
  other}\hspace{0.3cm}f_{\mu\nu\rho}=0.
  \label{sc-matrix}
\end{eqnarray}

When we expand the deformed I\hspace{-.1em}IB matrix model up to the
forth order of the quantum fluctuations, we obtain the following action:
\begin{eqnarray}
 \hat{S}
  &=&-{\rm Tr}\sum_{stu}\left(\frac14\left[p_i,p_j\right]^2
     -\frac{{\rm i}}{3}f_{ijk}[p^i,p^j]p^k\right. \nonumber \\
  &&+\,\frac12\left[p_i,a_{\mu}^{(s,t)}\right]^2
    -b^{(s,t)}\Bigl[p_i,[p^i,c^{(t,s)}]\Bigr]
    +\frac12\bar{\varphi}^{(s,t)}\Gamma^i\left[p_i,\varphi^{(t,s)}\right]
     \nonumber \\
  &&+\,\left[p_i,a_{\mu}^{(s,t)}\right]\!\!\left[a^{(t,u)i},a^{(u,s)\mu}\right]
     -b^{(s,t)}\Bigl[p_i,[a^{(t,u)i},c^{(u,s)}]\Bigr] \nonumber \\
  &&+\left.\frac12\bar{\varphi}^{(s,t)}\Gamma^{\mu}\left[a_{\mu}^{(t,u)},\varphi^{(u,s)}\right]
     -\frac{{\rm i}}{3}f_{ijk}\left[a^{(s,t)i},a^{(t,u)j}\right]a^{(u,s)k}
     +\frac14\left[a_{\mu}^{(s,t)},a_{\nu}^{(t,u)}\right]^2\right). \nonumber \\
 \label{hat-s}
\end{eqnarray}
Because of the $j_{st}=j_s-j_t$ depends only on $s-t$, we can impose
the following condition on the quantum fluctuations and ghosts and
anti-ghosts fields:
\begin{eqnarray}
 &&a_{\mu}^{(s+1,t+1)}=a_{\mu}^{(s,t)}, \hspace{0.5cm}
  \varphi^{(s+1,t+1)}=\varphi^{(s,t)}, \nonumber \\
 &&c^{(s+1,t+1)}=c^{(s,t)}, \hspace{0.5cm}
  b^{(s+1,t+1)}=b^{(s,t)}.
\end{eqnarray}
From the above condition, we obtain the condition for the mode expansion
coefficients:
\begin{eqnarray}
 &&a_{\mu JM}^{(s+1,t+1)}=a_{\mu JM}^{(s,t)}, \hspace{0.5cm}
  \varphi_{JM}^{(s+1,t+1)}=\varphi_{JM}^{(s,t)}, \nonumber \\
 &&c_{JM}^{(s+1,t+1)}=c_{JM}^{(s,t)}, \hspace{0.5cm}
  b_{JM}^{(s+1,t+1)}=b_{JM}^{(s,t)}.
\end{eqnarray}
We can rewrite this condition as follows:
\begin{eqnarray}
 &&a_{\mu JM}^{(s,t)}=a_{\mu JMj_{st}}, \hspace{0.5cm}
  \varphi_{JM}^{(s,t)}=\varphi_{JMj_{st}}, \nonumber \\
 &&c_{JM}^{(s,t)}=c_{JMj_{st}}, \hspace{0.5cm}
  b_{JM}^{(s,t)}=b_{JMj_{st}}.
  \label{mode}
\end{eqnarray}

We rewrite the action (\ref{hat-s}) by using (\ref{mode}).
For example, we consider the following term:
\begin{eqnarray}
 -{\rm Tr}
  \sum_{st}
  \left(\frac12
   \left[p_i,a^{(s,t)}_{\mu}\right]^2\right)
  &=&\frac{\beta^2}{2}
     {\rm Tr}
     \sum_{st}
     a^{(s,t)}_{\mu}
     L_i\circ L^i\circ
     \delta^{\mu\nu}
     a_{\nu}^{(t,s)}
     \nonumber \\
  &=&\frac{\beta^2}{2}{\rm Tr}
     \sum_{st}\sum_{J_1M_1}\sum_{J_2M_2}
     a^{(s,t)}_{\mu J_1M_1}\otimes\hat{Y}_{J_1M_1}^{(j_sj_t)}
     J_2\left(J_2+1\right)\delta^{\mu\nu}
     a_{\nu J_2M_2}^{(t,s)}\otimes\hat{Y}_{J_2M_2}^{(j_tj_s)}
     \nonumber \\
  &=&\frac{N_0\beta^2}{2}
     \sum_{st}
     \sum_{J=|j_s-j_t|}^{j_s+j_t}
     \sum_{M=-J}^{J}
      a^{(s,t)}_{\mu JM}
     \left(-1\right)^{-M-\left(j_s-j_t\right)}
     J\left(J+1\right)
     \delta^{\mu\nu}
     a^{(t,s)}_{\nu JM},
     \nonumber \\
\end{eqnarray}
where we use the property of fuzzy spherical harmonics in the appendix C.
We set that $s-t=p$ and $s=h$, so $p$ and $h$ are integers,
to make the connection between super
Yang-Mills theory on the $S^3$ and I\hspace{-.1em}IB matrix model
on the $S^3$.
In this way, we obtain that
\begin{eqnarray}
 \frac{N_0\beta^2}{2}
     \sum_{h}
     \sum_{J=0}^{\infty}
     \sum_{M=-J}^{J}
     \sum_{\tilde{M}=-J}^{J}
      a_{\mu JM\tilde{M}}
     \left(-1\right)^{-M-\tilde{M}}
     J\left(J+1\right)
     \delta^{\mu\nu}
     a_{\nu JM\tilde{M}},
\end{eqnarray}
where we set that $p/2=\tilde{M}$ and $a^{(s,t)}_{\mu JM}=a_{\mu
JM\tilde{M}}$. In a similar way, we can evaluate the other terms in
the action (\ref{hat-s}). Then, the overall factor $\sum_h$ appears
in front of the action. After factoring out the overall factor
$\sum_h$, we find that the mode expansion of the deformed
I\hspace{-.1em}IB matrix model around the background $Y_i$ is
identical to the action of super Yang-Mills on the $S^3$.

In what follows, we investigate the relations between the gauge
theory on the $S^3$ background and the matrix models on $S^3$.
Although they are classically equivalent, the relation is more
subtle at the quantum level since the matrix model compactification
assumes a definite cutoff procedure. We have evaluated the effective
action of the gauge theory on $S^3$ using the background field
method in section 3. In this section, we evaluate an effective
action $\hat{W}$ of the deformed I\hspace{-.1em}IB matrix model on
the $S^3$ background $Y_i$:
\begin{eqnarray}
 \hat{W}=-\log\!\int\!da\,d\varphi\,dc\,db\;{\rm e}^{-\hat{S}},
\end{eqnarray}
where $\hat{S}$ is the action (\ref{hat-s}). Firstly, we evaluate
the effective action at the tree level:
\begin{eqnarray}
 \hat{W}_{\rm tree}
  &=&-{\rm Tr}\left(\frac14\left[p_i,p_j\right]^2
     -\frac{{\rm i}}{3}f_{ijk}[p^i,p^j]p^k\right) \nonumber \\
  &=&-\frac{\beta^4}{6}{\rm Tr}\sum_{s=1}^{T}N_s\left(L_i^{[j_s]}\right)^2
   \nonumber \\
  &=&-\frac{\beta^4}{6}\sum_{s=1}^{T}N_sj_s\left(j_s+1\right)\left(2j_s+1\right).
\end{eqnarray}
We impose the conditions $2j_s+1=N_0+s$ in order to connect
super Yang-Mills theory on the $S^3$ and I\hspace{-.1em}IB matrix model
on the $S^3$.
We obtain
\begin{eqnarray}
 \hat{W}_{\rm tree}
  =-\frac{\beta^4}{24}
   \sum_{h=1}^{\infty}
   \left[\left(N_0+h\right)^3-\left(N_0+h\right)\right],
\end{eqnarray}
where we set that $s=h$, and $N_s=N=1$ for simplicity. Since $h$
takes integer values, the sum over $h$ is formally divergent. We
have a cutoff scale on $h$ at a number $2\Lambda=T$ which is equal
to the number of $(s,t)$ blocks. When we take a large $N_0$ limit in
such a way that $N_0\gg \Lambda$, we obtain the following effective
action at the tree level:
\begin{eqnarray}
 \hat{W}_{\rm tree}
  \mathop{\longrightarrow}_{{N_0\rightarrow\infty}\atop{N_0\gg \Lambda}}
  \sum_{h=1}^{2\Lambda}
  \left(-\frac{\beta^4}{24}N_0^3\right).
  \label{hat-w-tree}
\end{eqnarray}

Secondly, we calculate the effective action at the 1-loop level as follows:
\begin{eqnarray}
 \hat{W}_{\rm 1-loop}
  &\sim&-{\rm Tr}\sum_{st}\left(\frac{1}{P_i^2}\right)^2\!\!F_{ij}F^{ji}
    \nonumber \\
  &=&{\rm Tr}\sum_{st}\frac{2\beta^2}{P_i^2}
    \nonumber \\
  &=&\sum_{st}\sum_{J=|j_s-j_t|}^{j_s+j_t}\frac{2}{J\left(J+1\right)}\left(2J+1\right).
\end{eqnarray}
We obtain that
\begin{eqnarray}
 \hat{W}_{\rm 1-loop}
  \sim
   \sum_{h=1}^{\infty}
   \sum_{J=0}^{\infty}
   \sum_{\tilde{M}=-J}^{J}
  \frac{2}{J\left(J+1\right)}
  \left(2J+1\right).
\end{eqnarray}
We have a cutoff such that $h<2\Lambda$, so that the maximal value
of $J$ and $\tilde{M}$ are $N_0$ and $\Lambda$, respectively. We
divide the summation over $J$ into two parts at the value $\Lambda$
as the following:
\begin{eqnarray}
 \hat{W}_{\rm 1-loop}
  \sim
  \sum_{h=1}^{2\Lambda}
  \sum_{J=1/2}^{\Lambda}
  \frac{2}{J\left(J+1\right)}\left(2J+1\right)^2
  +\sum_{h=1}^{2\Lambda}
  \sum_{J=\Lambda+1/2}^{N_0}
  \frac{2}{J\left(J+1\right)}\left(2J+1\right)\left(2\Lambda+1\right),
\end{eqnarray}
where we omit a zero mode of $J$.
When we take a large $N_0$ limit with
$N_0\gg \Lambda$, we obtain the effective action at the 1-loop level:
\begin{eqnarray}
 \hat{W}_{\rm 1-loop}
  \mathop{\longrightarrow}_{{N_0\rightarrow\infty}\atop{N_0\gg \Lambda}}
  \sum_{h=1}^{2\Lambda}
  \left(8\Lambda+8\Lambda\log N_0\right).
\end{eqnarray}

Finally, we calculate the effective action at the 2-loop level due to
planar diagrams. We
describe the detailed calculations of the 2-loop effective action
in appendix D. The result is
\begin{eqnarray}
 \hat{W}_{\rm 2-loop}
  &\sim&
  \frac{36}{\beta^4N_0}
  \sum_{stu}
  \sum_{J_1=|j_s-j_t|}^{j_s+j_t}
  \sum_{J_2=|j_t-j_u|}^{j_t+j_u}
  \sum_{J_3=|j_u-j_s|}^{j_u+j_s}
   \nonumber \\
  &&\times
   \frac{\left(2J_1+1\right)\left(2J_2+1\right)\left(2J_3+1\right)}
        {J_1\left(J_1+1\right)J_2\left(J_2+1\right)J_3\left(J_3+1\right)}
  \left(
   \begin{array}{ccc}
        J_1     &      J_2    &    J_3 \\
    \tilde{M}_1 & \tilde{M}_2 & \tilde{M}_3 \\
   \end{array}
         \right)^2,
\end{eqnarray}
where $\tilde{M}_1=j_s-j_t$, $\tilde{M}_2=j_t-j_u$ and $\tilde{M}_3=j_u-j_s$.
We can further rewrite it as
\begin{eqnarray}
 \hat{W}_{\rm 2-loop}
  &\sim&
  \frac{36}{\beta^4N_0}
  \sum_{h=1}^{\infty}
  \sum_{J_1=1/2}^{\infty}
  \sum_{\tilde{M}_1=-J_1}^{J_1}
  \sum_{J_2=1/2}^{\infty}
  \sum_{\tilde{M}_2=-J_2}^{J_2}
  \sum_{J_3=1/2}^{\infty}
  \sum_{\tilde{M}_3=-J_3}^{J_3}
   \nonumber \\
  &&\times
   \frac{\left(2J_1+1\right)\left(2J_2+1\right)\left(2J_3+1\right)}
        {J_1\left(J_1+1\right)J_2\left(J_2+1\right)J_3\left(J_3+1\right)}
  \left(
   \begin{array}{ccc}
         J_1    &     J_2     &    J_3 \\
    \tilde{M}_1 & \tilde{M}_2 & \tilde{M}_3 \\
   \end{array}
         \right)^2,
\end{eqnarray}
where we set that $p/2=\tilde{M}_1$, $q/2=\tilde{M}_2$ and
$(-p-q)/2=\tilde{M}_3$, and omit a zero mode of $J_1$, $J_2$ and
$J_3$. We have the cutoff scale $2\Lambda$ on $h$, and divide the
summation over $J_1$, $J_2$ and $J_3$ into two sections at $\Lambda$
as following:
\begin{eqnarray}
 \hat{W}_{\rm 2-loop}
  &\sim&
  \frac{36}{\beta^4N_0}
  \sum_{h=1}^{2\Lambda}
  \sum_{J_1=1/2}^{\Lambda}
  \sum_{\tilde{M}_1=-J_1}^{J_1}
  \sum_{J_2=1/2}^{\Lambda}
  \sum_{\tilde{M}_2=-J_2}^{J_2}
  \sum_{J_3=1/2}^{\Lambda}
  \sum_{\tilde{M}_3=-J_3}^{J_3}
   \nonumber \\
  &&\times
   \frac{\left(2J_1+1\right)\left(2J_2+1\right)\left(2J_3+1\right)}
        {J_1\left(J_1+1\right)J_2\left(J_2+1\right)J_3\left(J_3+1\right)}
  \left(
   \begin{array}{ccc}
         J_1    &     J_2     &    J_3 \\
    \tilde{M}_1 & \tilde{M}_2 & \tilde{M}_3 \\
   \end{array}
         \right)^2 \nonumber \\
   &&+
    \frac{3\cdot36}{\beta^4N_0}
    \sum_{h=1}^{2\Lambda}
    \sum_{J_1=\Lambda+1/2}^{N_0}
    \sum_{J_2=1/2}^{\Lambda}
    \sum_{\tilde{M}_2=-J_2}^{J_2}
    \sum_{J_3=1/2}^{\Lambda}
    \sum_{\tilde{M}_3=-J_3}^{J_3}
   \nonumber \\
  &&\times\left(2\Lambda+1\right)
   \frac{\left(2J_1+1\right)\left(2J_2+1\right)\left(2J_3+1\right)}
  {J_1\left(J_1+1\right)J_2\left(J_2+1\right)J_3\left(J_3+1\right)}
  \left(
   \begin{array}{ccc}
         J_1    &     J_2     &    J_3 \\
        \Lambda & \tilde{M}_2 & \tilde{M}_3 \\
   \end{array}
         \right)^2 \nonumber \\
  &&+
   \frac{3\cdot36}{\beta^4N_0}
   \sum_{h=1}^{2\Lambda}
   \sum_{J_1=\Lambda+1/2}^{N_0}
   \sum_{J_2=\Lambda+1/2}^{N_0}
   \sum_{J_3=1/2}^{\Lambda}
   \sum_{\tilde{M}_3=-J_3}^{J_3}
   \nonumber \\
  &&\times
   \left(2\Lambda+1\right)^2
   \frac{\left(2J_1+1\right)\left(2J_2+1\right)\left(2J_3+1\right)}
        {J_1\left(J_1+1\right)J_2\left(J_2+1\right)J_3\left(J_3+1\right)}
  \left(
   \begin{array}{ccc}
         J_1    &     J_2     &    J_3 \\
      \Lambda   &   \Lambda   &  \tilde{M}_3  \\
   \end{array}
         \right)^2 \nonumber \\
  &&+
   \frac{36}{\beta^4N_0}
   \sum_{h=1}^{2\Lambda}
   \sum_{J_1=\Lambda+1/2}^{N_0}
   \sum_{J_2=\Lambda+1/2}^{N_0}
   \sum_{J_3=\Lambda+1/2}^{N_0}
   \nonumber \\
  &&\times
   \left(2\Lambda+1\right)^2
   \frac{\left(2J_1+1\right)\left(2J_2+1\right)\left(2J_3+1\right)}
        {J_1\left(J_1+1\right)J_2\left(J_2+1\right)J_3\left(J_3+1\right)}
  \left(
   \begin{array}{ccc}
         J_1    &     J_2     &    J_3 \\
        \Lambda &   \Lambda   &  \Lambda \\
   \end{array}
         \right)^2.
  \label{w-2loop-cutoff}
\end{eqnarray}
When we take the large $N_0$ limit with $N_0\gg \Lambda$, we find
that the first term of (\ref{w-2loop-cutoff}) is logarithmically
infinite while the others are finite. We describe the detailed
calculations of (\ref{w-2loop-cutoff}) in appendix D.9. In this way,
we obtain the effective action at the 2-loop level:
\begin{eqnarray}
 \hat{W}_{\rm 2-loop}
  \mathop{\longrightarrow}_{{N_0\rightarrow\infty}\atop{N_0\gg \Lambda}}
  \sum_{h=1}^{2\Lambda}
  \left(\frac{576\pi^2}{\beta^4}
  \frac{1}{N_0}\log \Lambda\right),
\end{eqnarray}
We summarize the effective action of the deformed I\hspace{-.1em}IB
matrix model on $S^3$ in a matrix model compactification procedure
up to 2-loop level:
\begin{eqnarray}
 \hat{w}
  \equiv
  \hat{W}/\sum_h
  \mathop{\longrightarrow}_{{N_0\rightarrow\infty}\atop{N_0\gg \Lambda}}
   -\frac{\beta^4}{24}N_0^3
   +8\Lambda+8\Lambda\log N_0
   +\frac{576\pi^2}{\beta^4}\frac{1}{N_0}\log\Lambda,
  \label{eff_action-matrix}
\end{eqnarray}
where we have factored out the overall factor $\sum_h$
in front of the effective action.

We make the comparison between the effective action
(\ref{eff_action-derivative}) of the gauge theory on $S^3$ and the effective action
(\ref{eff_action-matrix}) of a matrix model compactified on $S^3$.
Tr over matrices corresponds to the integration over the volume as
follows:
\begin{eqnarray}
 \frac{1}{N_0}{\rm Tr}
  \longrightarrow
  \int\!d\Omega.
\end{eqnarray}
Therefore, we obtain the following relation
by comparing the tree level action:
\begin{eqnarray}
 \alpha^4=\beta^4N_0.
\end{eqnarray}
as they are classically equivalent. We note here that the tree level
and the 1-loop contributions are highly divergent just like the
gauge theory on $S^3$ case in the preceding section. In fact they
are very sensitive to the cutoff procedure. However we find the
identical 2-loop contribution as it is only logarithmically
divergent. We also expect that the higher loop contributions are
finite since the 3-dimensional gauge theory is super renormalizable.
With our identification of the inverse coupling constant
$\alpha^4=\beta^4N_0$, we conclude again that the effective action
of the deformed I\hspace{-.1em}IB matrix model on the $S^3$ is
stable against the quantum corrections as it is dominated by the
tree level contribution.

%
%\newpage
%%%%%%%%%%%%%%%%%%%%%%%%%%%%%%%%%%%%%%%%%%%%%%%%%%%%%%%%%%%%%%%%%%%%%%%%%%%%%%%%%
%%%%%%%%%%%%%%%%%%%%%%%%%%%%%%%%%%%% conclusion %%%%%%%%%%%%%%%%%%%%%%%%%%%%%%%%%
%%%%%%%%%%%%%%%%%%%%%%%%%%%%%%%%%%%%%%%%%%%%%%%%%%%%%%%%%%%%%%%%%%%%%%%%%%%%%%%%%
\section{Conclusions and discussions}
In this paper, we have shown that the $S^3$ background is one of
nontrivial solutions of a deformed I\hspace{-.1em}IB matrix model.
 We have evaluated the effective actions of a deformed I\hspace{-.1em}IB
matrix model on the $S^3$ background up to the 2-loop level in a
two different cutoff procedure. We have found that the effective
action of the deformed I\hspace{-.1em}IB
matrix model on the $S^3$ background is stable on the
condition that the coupling constant is ${\cal O}(1)$.
Since we have evaluated only planar diagrams, our investigation is valid
in the large $N$ limit of $U(N)$ gauge theory on $S^3$.

In section 3, we have used the three derivatives ${\rm
i}\,\partial^{\,(1)}_i$ on the $S^3$ to evaluate the effective
action on the $S^3$ background.
In \cite{hkk}, the authors pointed
out that the bosonic parts $A_{\mu}$ of I\hspace{-.1em}IB matrix
model can be interpreted as derivatives on a curved space.
They claim that I\hspace{-.1em}IB matrix model represents generic curved
spaces in their interpretation. We believe that their claim is
clarified by our concrete investigations on $S^3$ especially at the
quantum level.

In \cite{istt}, the $S^3$ is realized by three matrices $Y_i$ which
are the vacuum configuration of a matrix model.
We have also investigated the effective actions on the $S^3$ background $Y_i$ in
this generalized matrix model compactification.
In both cases, we find that the highly divergent contributions at the
tree and 1-loop level are sensitive to the UV cutoff.
However the 2-loop level contributions are universal since they are only
logarithmically divergent.
We expect that the higher loop contributions are insensitive to the UV cutoff since 3-dimensional gauge theory
is super renormalizable.
We can thus conclude that the effective action of the deformed I\hspace{-.1em}IB matrix model on the $S^3$
is stable against the quantum corrections as it is dominated by the
tree level contribution.

We recall here that we have obtained the identical conclusions for
the $S^2$ case.
We thus believe that the 2- and 3-dimensional spheres
are classical objects in I\hspace{-.1em}IB matrix model since the tree level
effective action dominates.
We can in turn conclude that they are
not the solutions of the I\hspace{-.1em}IB matrix model without a Myers term.

We have investigated the I\hspace{-.1em}IB matrix model in a matrix model
compactification procedure by imposing the periodicity on the blocks
of the matrices.
We believe it is also interesting to investigate
the same background without such a condition.
Such a case appears to correspond to the quenched matrix model in the flat background case.
It is also interesting to investigate higher dimensional spaces such
as $S^4$ or $S^3\times R$ as they are physically and quantum
mechanically more interesting.

%%%%%%%%%%%%%%%%%%%%%%%%%%%%%%%%%%%%%%%%%%%%%%%%%%%%%%%%%%%%%%%%%%%%%%%%%%%%%%%%
%%%%%%%%%%%%%%%%%%%%%%%%%%%%%%%%% acknowledgments %%%%%%%%%%%%%%%%%%%%%%%%%%%%%%
%%%%%%%%%%%%%%%%%%%%%%%%%%%%%%%%%%%%%%%%%%%%%%%%%%%%%%%%%%%%%%%%%%%%%%%%%%%%%%%%
\section*{Acknowledgments}
This work is supported in part by the Grant-in-Aide for Scientific
Research from the Ministry of Education, Science and Culture of
Japan. We would like to thank S. Shimasaki for discussions during
the 1st Asian Winter School on ``String Theory, Geometry, Holography
and Black Holes''.
%
%\newpage
%%%%%%%%%%%%%%%%%%%%%%%%%%%%%%%%%%%%%%%%%%%%%%%%%%%%%%%%%%%%%%%%%%%%%%%%%%%%%%%%
%%%%%%%%%%%%%%%%%%%%%%%%%%%%%%%%%%%%% appendix %%%%%%%%%%%%%%%%%%%%%%%%%%%%%%%%%
%%%%%%%%%%%%%%%%%%%%%%%%%%%%%%%%%%%%%%%%%%%%%%%%%%%%%%%%%%%%%%%%%%%%%%%%%%%%%%%%
\appendix
\section{Spherical harmonics on $S^3$}
In this appendix, we summarize the spherical harmonics on the $S^3$.
We adopt the following representation of the spherical harmonics on the
$S^3$ \cite{br}:
\begin{eqnarray}
 Y^{\;n}_{m\tilde{m}}\left(\Omega\right)
  =\sqrt{n+1}\,
   {\rm e}^{{\rm i}m\phi+{\rm i}\tilde{m}\tilde{\phi}}\,
   d^{\frac{n}{2}}_{\frac12\left(m+\tilde{m}\right),
                    \frac12\left(m-\tilde{m}\right)}\left(2\theta\right),
\end{eqnarray}
where
\begin{eqnarray}
 d^J_{M,\tilde{M}}\left(2\theta\right)
  &=&\left[\frac{(J+M)!(J-M)!}{(J+\tilde{M})!(J-\tilde{M})!}\right]^{1/2}
   \nonumber \\
  &&\times\left(\cos\theta\right)^{M+\tilde{M}}
          \left(\sin\theta\right)^{M-\tilde{M}}
          P^{(M-\tilde{M},\,M+\tilde{M})}_{J-M}\left(\cos 2\theta\right),
\end{eqnarray}
$J=n/2=0,1/2,1,\cdots$, $M=\left(m+\tilde{m}\right)/2=-J,-J+1,\cdots,J-1,J$ and
$\tilde{M}=\left(m-\tilde{m}\right)/2=-J,-J+1,\cdots,J-1,J$.
Additionally, we make use of the Rodrigues formulas for the Jacobi
polynomial:
\begin{eqnarray}
 P^{(M-\tilde{M},\,M+\tilde{M})}_{J-M}\left(\cos 2\theta\right)
  &=&\frac{(-1)^{J-M}}{2^{J-M}\left(J-M\right)!}
     \left(1-\cos 2\theta\right)^{-M+\tilde{M}}
     \left(1+\cos 2\theta\right)^{-M-\tilde{M}}
     \nonumber \\
  &&\times\frac{d^{J-M}}{d\cos 2\theta^{J-M}}
    \left[\left(1-\cos 2\theta\right)^{J-\tilde{M}}
          \left(1+\cos 2\theta\right)^{J+\tilde{M}}\right].
\end{eqnarray}
$Y^{\,n}_{m\tilde{m}}$ satisfies the following equations:
\begin{eqnarray}
 &&\mbox{\boldmath $J$}^2_1\,Y^{\;n}_{m\tilde{m}}\left(\Omega\right)
  =\frac{n}{2}\left(\frac{n}{2}+1\right)Y^{\;n}_{m\tilde{m}}\left(\Omega\right),
  \nonumber \\
 &&\mbox{\boldmath $J$}^2_2\,Y^{\;n}_{m\tilde{m}}\left(\Omega\right)
  =\frac{n}{2}\left(\frac{n}{2}+1\right)Y^{\;n}_{m\tilde{m}}\left(\Omega\right),
  \nonumber \\
 &&J_{13}\,Y^{\;n}_{m\tilde{m}}\left(\Omega\right)
  =\frac12\left(m+\tilde{m}\right)Y^{\;n}_{m\tilde{m}}\left(\Omega\right),
  \nonumber \\
 &&J_{23}\,Y^{\;n}_{m\tilde{m}}\left(\Omega\right)
  =\frac12\left(m-\tilde{m}\right)Y^{\;n}_{m\tilde{m}}\left(\Omega\right),
\end{eqnarray}
where $J_{1i}$ and $J_{2i}$ are the generators of $SU(2)$ algebra.
$Y^{\,n}_{m\tilde{m}}$ are normalized as follows:
\begin{eqnarray}
 \int\!d\Omega\,
  Y^{\;n_1}_{m_1\tilde{m}_1}\left(\Omega\right)\,
  Y^{\;n_2}_{m_2\tilde{m}_2}\left(\Omega\right)
  =(-1)^{-\tilde{m}_1}\,
   \delta_{n_1n_2}\,
   \delta_{m_1-m_2}\,
   \delta_{\tilde{m}_1-\tilde{m}_2},
\end{eqnarray}
The complex conjugate of $Y^{\;n}_{m\tilde{m}}$ is that
\begin{eqnarray}
 Y^{\;n\;\ast}_{m\tilde{m}}\left(\Omega\right)
  =(-1)^{\tilde{m}}Y^{\;\;\;\;n}_{-m-\tilde{m}}\left(\Omega\right).
\end{eqnarray}
The integrals of the product of three spherical harmonics \cite{cutkosky,sen} can be
evaluated as
\begin{eqnarray}
 \int\!d\Omega\,
  Y^{\;n_1}_{m_1\tilde{m}_1}\left(\Omega\right)\,
  Y^{\;n_2}_{m_2\tilde{m}_2}\left(\Omega\right)\,
  Y^{\;n_3}_{m_3\tilde{m}_3}\left(\Omega\right)
  &=&(-1)^{(n_1+n_2+n_3)/2}\sqrt{(n_1+1)(n_2+1)(n_3+1)} \nonumber \\
  &&\times\left(
         \begin{array}{ccc}
          \frac{n_1}{2} & \frac{n_2}{2} & \frac{n_3}{2} \\
                        &               &               \\
          M_1           & M_2           & M_3 \\
         \end{array}
         \right)\!
         \left(
         \begin{array}{ccc}
          \frac{n_1}{2} & \frac{n_2}{2} & \frac{n_3}{2} \\
                        &               &              \\
          \tilde{M_1}   & \tilde{M_2}   & \tilde{M_3} \\
         \end{array}
         \right), \nonumber \\
 \label{3-spherical_harmonics}
\end{eqnarray}
where $(\cdots)$ represents the 3-$j$ symbol of Wigner \cite{edmonds,vmk}.
\section{Two-loop effective action on background ${\rm i}\,\partial^{\,(1)}_i$}
In this appendix, we evaluate the effective action at the 2-loop
level on the background: ${\rm i}\,\partial^{\,(1)}_i$. We can
evaluate the effective action $W$ of the deformed I\hspace{-.1em}IB
matrix model on the $S^3$ as follows:
\begin{eqnarray}
 W&=&-\log\!\int\!da\,d\varphi\,dc\,db\;
    {\rm e}^{-\tilde{S}} \nonumber \\
  &=&W_{\rm tree}+W_{\rm 1-loop}+W_{\rm 2-loop},
\end{eqnarray}
where
\begin{eqnarray}
 W_{\rm 2-loop}
  &=&\left<\exp\left\{{\rm Tr}\left([p_i, a_{\mu}][a^i,a^{\mu}]
            -b\Bigl[p_i,[a^i, c]\Bigr]\right.\right.\right. \nonumber \\
  &&\left.\left.\left.
             +\,\frac12\bar{\varphi}\Gamma^{\mu}[a_{\mu},\varphi]
             -\frac{{\rm i}}{3}f_{ijk}[a^i,a^j]a^k
             +\frac14\left[a_{\mu},a_{\nu}\right]^2\right)\right\}\right>_{\rm 1PI},
\end{eqnarray}
and $\left<\cdots\right>_{\rm 1PI}$ represents that we sum only the
1PI (1-Particle-Irreducible) diagrams.
There are five 1PI diagrams to evaluate which are illustrated in Fig. 2.
The diagrams (a), (b) and (c) represent the contributions from gauge fields, and (c)
involves the Myers type interaction.
The diagrams (d) and (e) represent the contributions from ghost and fermion fields respectively.
\begin{figure}[htbp]
 \begin{center}
  \includegraphics[scale=1.0]{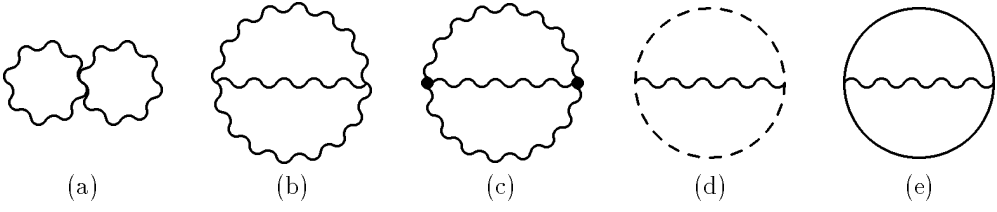}
  \caption{Feynman diagrams of 2-loop corrections to the effective
  action \cite{iktt-2004}.}
  \label{fig:feynman_diagrams}
 \end{center}
\end{figure}
\subsection{Bosonic propagators}
From the quadratic terms for the gauge fields $a_{\mu}$, we can read out
the propagators of gauge boson modes $a^{\;\;\;n}_{\mu m\tilde{m}}$.
\begin{eqnarray}
 &&{\rm Tr}\left(\frac12 a_{\mu}P_i^2\delta^{\mu\nu}a_{\nu}\right)
  \nonumber \\
 &\longrightarrow&
   \int\!d\Omega
   \left(\frac12\sum_{n_1m_1\tilde{m}_1}\!\sum_{n_2m_2\tilde{m}_2}\!\!\!
   a^{\;\;\;n_1}_{\mu m_1\tilde{m}_1}
   Y^{\;n_1}_{m_1\tilde{m}_1}\left(\Omega\right)\,
   \frac{\alpha^2}{4}n_2(n_2+2)\,
   \delta^{\mu\nu}\,
   a^{\;\;\;n_2}_{\nu m_2\tilde{m}_2}
   Y^{\;n_2}_{m_2\tilde{m}_2}\left(\Omega\right)\right) \nonumber \\
 &=&\frac12\!\sum_{n_1m_1\tilde{m}_1}\sum_{n_2m_2\tilde{m}_2}\!\!\!
   a^{\;\;\;n_1}_{\mu m_1\tilde{m}_1}
   \frac{\alpha^2}{4}n_2(n_2+2)\,
   \delta^{\mu\nu}(-1)^{-\tilde{m}_1}
   \delta_{n_1n_2}\,
   \delta_{m_1-m_2}\,
   \delta_{\tilde{m}_1-\tilde{m}_2}\,
   a^{\;\;\;n_2}_{\nu m_2\tilde{m}_2}. \nonumber \\
\end{eqnarray}
In the first step, we have taken the semi-classical limit and
substituted the quantum fluctuations which are expanded by the
spherical harmonics on the $S^3$:
\begin{eqnarray}
 a_{\mu}=\sum_{nm\tilde{m}}
  a^{\;\;\;n}_{\mu m\tilde{m}}\,
  Y^{\;n}_{m\tilde{m}}\left(\Omega\right).
\end{eqnarray}
Therefore,
\begin{eqnarray}
 \left<a^{\;\;\;n_1}_{\mu m_1\tilde{m}_1}
  a^{\;\;\;n_2}_{\nu m_2\tilde{m}_2}\right>
  =\frac{4}{\alpha^2}\frac{(-1)^{\tilde{m}_1}}{n_1(n_1+2)}\,
   \delta_{\mu\nu}\,
   \delta_{n_1n_2}\,
   \delta_{m_1-m_2}\,
   \delta_{\tilde{m}_1-\tilde{m}_2}.
\end{eqnarray}
The propagators of gauge boson fields become
\begin{eqnarray}
 \left<a_{\mu}\,a_{\nu}\right>
  =\frac{4}{\alpha^2}\sum_{nm\tilde{m}}
  \frac{(-1)^{\tilde{m}}}{n(n+2)}\,
  \delta_{\mu\nu}\,
  Y^{\;n}_{m\tilde{m}}\left(\Omega_1\right)\,
  Y^{\;\;\;\;n}_{-m-\tilde{m}}\left(\Omega_2\right).
\end{eqnarray}
In the same way, we can read off the propagators of ghost fields.
\begin{eqnarray}
 &&\left<c^{\;\;n_1}_{m_1\tilde{m}_1}b^{\;\;n_2}_{m_2\tilde{m}_2}\right>
  =\frac{4}{\alpha^2}\frac{(-1)^{\tilde{m}_1}}{n_1(n_1+2)}\,
   \delta_{n_1n_2}\,
   \delta_{m_1-m_2}\,
   \delta_{\tilde{m}_1-\tilde{m}_2}, \nonumber \\
 &&\left<c\,b\right>
  =\frac{4}{\alpha^2}\sum_{nm\tilde{m}}\frac{(-1)^{\tilde{m}}}{n(n+2)}\,
    Y^{\;n}_{m\tilde{m}}\left(\Omega_1\right)\,
    Y^{\;\;\;\;n}_{-m-\tilde{m}}\left(\Omega_2\right).
\end{eqnarray}
We have introduced the quantum fluctuations which are expanded by
the spherical harmonics on the $S^3$ as follows:
\begin{eqnarray}
 c&=&\sum_{nm\tilde{m}}c^{\;\;n}_{m\tilde{m}}\,
    Y^{\;n}_{m\tilde{m}}\left(\Omega\right), \nonumber \\
 b&=&\sum_{nm\tilde{m}}b^{\;\;n}_{m\tilde{m}}\,
    Y^{\;n}_{m\tilde{m}}\left(\Omega\right).
\end{eqnarray}
\subsection{Contribution from four-point gauge boson vertex (a)}
We evaluate the 1PI diagram involving a 4-point gauge boson vertex.
\begin{eqnarray}
 V_4&=&\frac14{\rm Tr}\left[a_{\mu},a_{\nu}\right]^2 \nonumber \\
    &=&\frac12{\rm Tr}\left(a_{\mu}a_{\nu}a^{\mu}a^{\nu}
                      -a_{\mu}a^{\mu}a_{\nu}a^{\nu}\right).
\end{eqnarray}
We can calculate $V_4$ using the Wick contraction.
\begin{eqnarray}
 V_4
  &\longrightarrow&
   \left[\frac12\left(10+10\right)-\frac12\left(10^2+10\right)\right]
   \int\!d\Omega_1\left(\frac{4}{\alpha^2}\right)^2\!\!\!\!
   \sum_{n_1m_1\tilde{m}_1}\sum_{n_2m_2\tilde{m}_2}
   \frac{(-1)^{\tilde{m}_1+\tilde{m}_2}}{n_1(n_1+2)n_2(n_2+2)} \nonumber \\
  &&\times
   Y^{\;n_1}_{m_1\tilde{m}_1}\left(\Omega_1\right)\,
   Y^{\;\;\;\;n_1}_{-m_1-\tilde{m}_1}\left(\Omega_1\right)\,
   Y^{\;n_2}_{m_2\tilde{m}_2}\left(\Omega_1\right)\,
   Y^{\;\;\;\;n_2}_{-m_2-\tilde{m}_2}\left(\Omega_1\right) \nonumber \\
  &=&-45\int\!d\Omega_1\!\int\!d\Omega_2
   \left(\frac{4}{\alpha^2}\right)^2\!\!\!\!
   \sum_{n_1m_1\tilde{m}_1}\sum_{n_2m_2\tilde{m}_2}
   \frac{(-1)^{\tilde{m}_1+\tilde{m}_2}}{n_1(n_1+2)n_2(n_2+2)} \nonumber \\
  &&\times
   Y^{\;n_1}_{m_1\tilde{m}_1}\left(\Omega_1\right)\,
   Y^{\;\;\;\;n_1}_{-m_1-\tilde{m}_1}\left(\Omega_2\right)\,
   Y^{\;n_2}_{m_2\tilde{m}_2}\left(\Omega_1\right)\,
   Y^{\;\;\;\;n_2}_{-m_2-\tilde{m}_2}\left(\Omega_2\right)\,
   \delta\left(\Omega_1-\Omega_2\right).
\end{eqnarray}
Here we can insert the complete set as follows:
\begin{eqnarray}
 \sum_{nm\tilde{m}}(-1)^{\tilde{m}}\,
  Y^{\;n}_{m\tilde{m}}\left(\Omega_1\right)\,
  Y^{\;\;\;\;n}_{-m-\tilde{m}}\left(\Omega_2\right)
  =\delta\left(\Omega_1-\Omega_2\right).
\end{eqnarray}
Therefore, we can get
\begin{eqnarray}
 V_4
  \longrightarrow
  -45\sum_{123}
  \Psi_{123}^{\ast}\left(\Omega_2\right)
  \frac{1}{P^2Q^2}
  \Psi_{123}\left(\Omega_1\right),
\end{eqnarray}
where $P$, $Q$ and $R$ are defined as follows:
\begin{eqnarray}
 &&P_iY^{\;n_1}_{m_1\tilde{m}_1}(\Omega)
  \equiv
  p_iY^{\;n_1}_{m_1\tilde{m}_1}(\Omega),
  \nonumber \\
 &&Q_iY^{\;n_2}_{m_2\tilde{m}_2}(\Omega)
  \equiv
  p_iY^{\;n_2}_{m_2\tilde{m}_2}(\Omega),
  \nonumber \\
 &&R_iY^{\;n_3}_{m_3\tilde{m}_3}(\Omega)
  \equiv
  p_iY^{\;n_3}_{m_3\tilde{m}_3}(\Omega).
\end{eqnarray}
We have introduced the wave functions such that
\begin{eqnarray}
 \Psi_{123}\left(\Omega\right)
  \equiv \int\!d\Omega\,
         Y^{\;n_1}_{m_1\tilde{m}_1}\left(\Omega\right)
         Y^{\;n_2}_{m_2\tilde{m}_2}\left(\Omega\right)
         Y^{\;n_3}_{m_3\tilde{m}_3}\left(\Omega\right).
\end{eqnarray}
$\sum_{123}$ denotes
$\sum_{n_1m_1\tilde{m}_1}\sum_{n_2m_2\tilde{m}_2}\sum_{n_3m_3\tilde{m}_3}$.
\subsection{Contribution from three-point gauge boson vertex (b)}
We evaluate the 1PI diagram involving 3-point gauge boson
vertices. We can express the contribution corresponding the diagram
(b) as follows:
\begin{eqnarray}
 V_3=\frac12\left({\rm Tr}[p_i,a_{\mu}][a^i, a^{\mu}]\right)^2.
\end{eqnarray}
We can express the result as a compact form:
\begin{eqnarray}
 V_3
  \longrightarrow
  \frac92
  \sum_{123}
  \Psi_{123}^{\ast}\left(\Omega_2\right)\,
  \frac{2P^2-P\cdot Q-P\cdot R}{P^2Q^2R^2}\,
  \Psi_{123}\left(\Omega_1\right).
\end{eqnarray}
We use the following relation:
\begin{eqnarray}
 P\cdot Q\,\Psi_{123}\left(\Omega\right)
  =\frac{R^2-P^2-Q^2}{2}\,
  \Psi_{123}\left(\Omega\right),
\end{eqnarray}
and the momentum conserved relation: $P+Q+R=0$. Therefore, we can
simplify the result as
\begin{eqnarray}
 V_3
  \longrightarrow
   \frac{27}{3}
   \sum_{123}\Psi_{123}^{\ast}\left(\Omega_2\right)
   \frac{1}{P^2Q^2}\Psi_{123}\left(\Omega_1\right).
\end{eqnarray}
We note the relation that
\begin{eqnarray}
 \sum_{m\tilde{m}}\,(-1)^{\tilde{m}}
  Y^{\;n}_{m\tilde{m}}\left(\Omega\right)
  P_i\,Y^{\;\;\;\;n}_{-m-\tilde{m}}\left(\Omega\right)
  =-\sum_{m\tilde{m}}\,(-1)^{\tilde{m}}
  \left(P_i\,Y^{\;n}_{m\tilde{m}}\left(\Omega\right)\right)
  Y^{\;\;\;\;n}_{-m-\tilde{m}}\left(\Omega\right).
\end{eqnarray}
\subsection{Contribution from the Myers type interaction (c)}
The diagram involving the Myers type interactions is represented as
follows:
\begin{eqnarray}
 V_{\rm M}=
  -\frac{1}{18}
  \left({\rm Tr}\,f_{ijk}[a^i,a^j]a^k\right)^2.
\end{eqnarray}
We get the following result
\begin{eqnarray}
 V_{\rm M}
  \longrightarrow
   4\alpha^2
   \sum_{123}
   \Psi_{123}^{\ast}\left(\Omega_2\right)\,
   \frac{1}{P^2Q^2R^2}\,
   \Psi_{123}\left(\Omega_1\right),
\end{eqnarray}
where we have used the relation: $f_{ijk}f^{ijk}=6\alpha^2$.
\subsection{Contribution from the ghost interaction (d)}
We evaluate the contribution from the ghost interactions.
\begin{eqnarray}
 V_{\rm gh}=\frac12\left({\rm Tr}\,[p_i,b][a^i,c]\right)^2.
\end{eqnarray}
The result is
\begin{eqnarray}
 V_{\rm gh}
  \longrightarrow
   -\frac12
   \sum_{123}
   \Psi_{123}^{\ast}\left(\Omega_2\right)\,
   \frac{1}{P^2Q^2}\,
   \Psi_{123}\left(\Omega_1\right).
\end{eqnarray}
\subsection{Fermion propagator}
We can read off the fermion propagator from the quadratic terms in
the action (\ref{tilde-s}):
\begin{eqnarray}
 &&{\rm Tr}\left(-\frac12\,\bar{\varphi}\Gamma^iP_i\varphi\right)
   \nonumber \\
 &\longrightarrow&
   \int\!d\Omega
   \left(-\frac12\!
    \sum_{n_1m_1\tilde{m}_1}
    \sum_{n_2m_2\tilde{m}_2}\!\!\!
   \bar{\varphi}^{\;\;n_1}_{m_1\tilde{m}_1}
   Y^{\;n_1\;\ast}_{m_1\tilde{m}_1}\left(\Omega\right)
   \Gamma^iP_i\,
   \varphi^{\;\;n_2}_{m_2\tilde{m}_2}
   Y^{\;n_2}_{m_2\tilde{m}_2}\left(\Omega\right)
  \right) \nonumber \\
 &=&-\frac12
   \sum_{n_1m_1\tilde{m}_1}
   \sum_{n_2m_2\tilde{m}_2}\!\!\!
   \bar{\varphi}^{\;\;n_1}_{m_1\tilde{m}_1}
   \Gamma^iP_i\,
   \delta_{n_1n_2}\,
   \delta_{m_1m_2}\,
   \delta_{\tilde{m}_1\tilde{m}_2}\,
   \varphi^{\;\;n_2}_{m_2\tilde{m}_2},
\end{eqnarray}
where we have expanded the quantum fluctuations by the spherical
harmonics on the $S^3$ as follows:
\begin{eqnarray}
 \varphi=\sum_{nm\tilde{m}}
       \varphi^{\;\;n}_{m\tilde{m}}
       Y^{\;n}_{m\tilde{m}}\left(\Omega\right).
\end{eqnarray}
Therefore,
\begin{eqnarray}
 \left<\varphi^{\;\;n_1}_{m_1\tilde{m}_1}
  \bar{\varphi}^{\;\;n_2}_{m_2\tilde{m}_2}\right>
 =-\frac{1}{\Gamma^iP_i}\,
   \delta_{n_1n_2}\,
   \delta_{m_1m_2}\,
   \delta_{\tilde{m}_1\tilde{m}_2}.
\end{eqnarray}
The fermion propagator is
\begin{eqnarray}
 \left<\varphi\,\bar{\varphi}\right>
 =\sum_{nm\tilde{m}}
  \left(-\frac{1}{\Gamma^iP_i}\right)
   (-1)^{\tilde{m}}
  Y^{\;n}_{m\tilde{m}}\left(\Omega_1\right)
  Y^{\;\;\;\;n}_{-m-\tilde{m}}\left(\Omega_2\right).
\end{eqnarray}
We can further expand the fermion propagator in powers of $1/P^2$ as
follows:
\begin{eqnarray}
 -\frac{1}{\Gamma^iP_i}
  &=&-\frac{1}{P^2+\frac{{\rm i}}{2}F_{ij}\Gamma^{ij}}\,
     \Gamma^kP_k \nonumber \\
  &=&-\frac{1}{P^2}\,\Gamma^iP_i
     +\frac{{\rm i}}{2}\left(\frac{1}{P^2}\right)^2\!\!
     F_{ij}\Gamma^i\Gamma^j\Gamma^kP_k
     +\frac14\left(\frac{1}{P^2}\right)^3\!\!
     F_{ij}\Gamma^{ij}F_{kl}\Gamma^{kl}\Gamma^aP_a
     +\cdots. \nonumber \\
 \label{fermion_propagator-exp}
\end{eqnarray}
The second term of (\ref{fermion_propagator-exp}) is that
\begin{eqnarray}
 \frac{{\rm i}}{2}\left(\frac{1}{P^2}\right)^2\!\!
  F_{ij}\Gamma^i\Gamma^j\Gamma^kP_k
  &=&\frac{{\rm i}}{2}\left(\frac{1}{P^2}\right)^2\!\!
     \left(f_{ijk}\Gamma^{ijl}P^kP_l
                 -2{\rm i}\,\Gamma\cdot P
                  +\cdots\right) \nonumber \\
  &=&\frac{{\rm i}}{2}\left(\frac{1}{P^2}\right)^2\!\!
     f_{ijk}\Gamma^{ijl}P^kP_l
     +\left(\frac{1}{P^2}\right)^2\!\!\Gamma\cdot P
     +\cdots.
\end{eqnarray}
Here we have used the formula as follows:
\begin{eqnarray}
 \Gamma^i\Gamma^j\Gamma^k
  =\Gamma^{ijk}
   +\delta^{ij}\Gamma^k
   -\delta^{ik}\Gamma^j
   +\delta^{jk}\Gamma^i.
\end{eqnarray}
While the third term of (\ref{fermion_propagator-exp}) is
\begin{eqnarray}
 \frac14\left(\frac{1}{P^2}\right)^3F_{ij}\Gamma^{ij}F_{kl}\Gamma^{kl}\Gamma^aP_a
  &=&\frac14\left(\frac{1}{P^2}\right)^3\left(-4\Gamma\cdot PP^2
                                        +\cdots\right) \nonumber \\
  &=&-\left(\frac{1}{P^2}\right)^2\Gamma\cdot P+\cdots.
\end{eqnarray}
Here we have used the formula as follows:
\begin{eqnarray}
 \Gamma^{ij}\Gamma^{kl}\Gamma^a
  &=&\Gamma^{ijkla}
     -\delta^{ik}\Gamma^{jla}
     +\delta^{il}\Gamma^{jka}
     -\delta^{ia}\Gamma^{jkl}
     +\delta^{jk}\Gamma^{ila}
     -\delta^{jl}\Gamma^{ika}
     +\delta^{ja}\Gamma^{ikl} \nonumber \\
   &&-\delta^{ka}\Gamma^{ijl}
     +\delta^{la}\Gamma^{ijk}
     -\delta^{ik}\delta^{jl}\Gamma^a
     +\delta^{ik}\delta^{ja}\Gamma^l
     +\delta^{il}\delta^{jl}\Gamma^k
     -\delta^{il}\delta^{ja}\Gamma^k \nonumber \\
   &&+\delta^{ia}\delta^{jk}\Gamma^l
     -\delta^{ia}\delta^{jl}\Gamma^k
     +\delta^{ka}\delta^{il}\Gamma^j
     -\delta^{ka}\delta^{jl}\Gamma^i
     -\delta^{la}\delta^{ik}\Gamma^j
     +\delta^{la}\delta^{jk}\Gamma^i. \nonumber \\
\end{eqnarray}
In this way, we obtain
\begin{eqnarray}
 -\frac{1}{\Gamma^iP_i}
  =-\frac{1}{P^2}\,\Gamma^iP_i
   +\frac{{\rm i}}{2}\left(\frac{1}{P^2}\right)^2\!\!f_{ijk}\Gamma^{ijl}P^{k}P_{l}
   +{\cal O}\left(\left(\frac{1}{P^2}\right)^3\right).
\end{eqnarray}
\subsection{Contribution from the fermion interaction (e)}
Finally, we evaluate the diagram involving fermion interactions.
\begin{eqnarray}
 V_{\rm F}
  &=&\frac18\left({\rm
             Tr}\,\bar{\varphi}\Gamma^{\mu}[a_{\mu},\varphi]\right)^2
      \nonumber \\
  &=&\frac12\left({\rm Tr}\,\bar{\varphi}\Gamma^{\mu}a_{\mu}\varphi\right)^2.
\end{eqnarray}
We perform the Wick contractions and evaluate it in the
semi-classical limit:
\begin{eqnarray}
 V_{\rm F}
  &\longrightarrow&
   \frac12\int\!d\Omega_1\!\int\!d\Omega_2 \nonumber \\
  &&\times{\rm tr}\!\!\!
    \sum_{n_1m_1\tilde{m}_1}\left(\frac{1}{P^2}\,\Gamma^iP_i
     -\frac{{\rm i}}{2}\left(\frac{1}{P^2}\right)^2\!\!f_{ijk}\Gamma^{ijl}P^kP_l
     +\cdots\right)\!\!
     \left(\frac{1+\Gamma_{11}}{2}\right)\Gamma^{\mu} \nonumber \\
  &&\times (-1)^{\tilde{m}_1}
    Y^{\;n_1}_{m_1\tilde{m}_1}\left(\Omega_1\right)
    Y^{\;\;\;\;n_1}_{-m_1-\tilde{m}_1}\left(\Omega_2\right) \nonumber \\
  &&\times\sum_{n_2m_2\tilde{m}_2}
    \frac{4}{\alpha^2}\frac{(-1)^{\tilde{m}_2}}{n_2(n_2+2)}\,
    \delta_{\mu\nu}\,
    Y^{\;n_2}_{m_2\tilde{m}_2}\left(\Omega_1\right)
    Y^{\;\;\;\;n_2}_{-m_2-\tilde{m}_2}\left(\Omega_2\right) \nonumber \\
  &&\times\sum_{n_3m_3\tilde{m}_3}
    \left(-\frac{1}{P^2}\,\Gamma^aP_a
    +\frac{{\rm i}}{2}\left(\frac{1}{P^2}\right)^2\!\!f_{abc}\Gamma^{abd}P^cP_d
    +\cdots\right)\!\!
    \left(\frac{1+\Gamma_{11}}{2}\right)\Gamma^{\nu} \nonumber \\
  &&\times (-1)^{\tilde{m}_3}
    Y^{\;n_3}_{m_3\tilde{m}_3}\left(\Omega_1\right)
    Y^{\;\;\;\;n_3}_{-m_3-\tilde{m}_3}\left(\Omega_2\right), \nonumber \\
\end{eqnarray}
where tr represents the trace over gamma matrices. Firstly, we
evaluate the leading term in the $1/P^2$ expansion. The trace of
products of gamma matrices in the leading term is as evaluated
follows:
\begin{eqnarray}
 {\rm tr}\Gamma^i\Gamma^{\mu}\Gamma^a\Gamma_{\mu}\!
  \left(\frac{1+\Gamma_{11}}{2}\right)
  &=&-8{\rm tr}\Gamma^i\Gamma^a \!
       \left(\frac{1+\Gamma_{11}}{2}\right) \nonumber \\
  &=&-128\,\delta^{ia}.
\end{eqnarray}
The trace of products of gamma matrices in the next leading term is
evaluated as follows:
\begin{eqnarray}
 {\rm tr}\Gamma^i\Gamma^{\mu}\Gamma^{abd}\Gamma_{\mu}\!
  \left(\frac{1+\Gamma_{11}}{2}\right)
  &=&-8{\rm tr}\Gamma^i\Gamma^{abd}\!
     \left(\frac{1+\Gamma_{11}}{2}\right) \nonumber \\
  &=&0.
\end{eqnarray}
The trace of products of gamma matrices in the next next leading
term is evaluated as follows:
\begin{eqnarray}
 {\rm tr}\Gamma^{ijl}\Gamma^{\mu}\Gamma^{abd}\Gamma_{\mu}\!
   \left(\frac{1+\Gamma_{11}}{2}\right)
  &=&4{\rm tr}\Gamma^{ijl}\Gamma^{adb}\!
    \left(\frac{1+\Gamma_{11}}{2}\right) \nonumber \\
  &=&-64
    \left(\delta^{ia}\delta^{jd}\delta^{lb}
    -\delta^{ia}\delta^{jb}\delta^{ld}
    -\delta^{ja}\delta^{id}\delta^{lb}\right. \nonumber \\
  &&\left.+\delta^{ja}\delta^{ib}\delta^{ld}
    +\delta^{la}\delta^{id}\delta^{jb}
    -\delta^{la}\delta^{ib}\delta^{jd}\right).
\end{eqnarray}
In this way, we obtain the following result
\begin{eqnarray}
 V_{\rm F}
  \sim
   \sum_{123}
   \Psi_{123}^{\ast}\left(\Omega_2\right)
   \left[-64\,\frac{P\cdot Q}{P^2Q^2R^2}
    +32\alpha^2\,\frac{P^2Q^2}{\left(P^2\right)^2\left(Q^2\right)^2R^2}
   \right]
   \Psi_{123}\left(\Omega_1\right). \nonumber \\
\end{eqnarray}
\subsection{Two-loop effective action}
We summarize the 2-loop effective action on the $S^3$ as follows:
\begin{eqnarray}
 W_{\rm 2-loop}
  &\sim&V_4+V_3+V_{\rm M}+V_{\rm gh}+V_{\rm F} \nonumber \\
  &=&36\alpha^2
     \sum_{123}
     \Psi_{123}^{\ast}\left(\Omega_2\right)\,
     \frac{1}{P^2Q^2R^2}\,
     \Psi_{123}\left(\Omega_1\right) \nonumber \\
  &=&\frac{2304}{\alpha^4}\sum_{n_1n_2n_3}
     \frac{(n_1+1)(n_2+1)(n_3+1)}{n_1(n_1+2)n_2(n_2+2)n_3(n_3+2)}.
\end{eqnarray}
Here we have used the formula (\ref{3-spherical_harmonics}), and
evaluated the summations over $m_1$, $m_2$, $m_3$, $\tilde{m}_1$,
$\tilde{m}_2$ and $\tilde{m}_3$.
\section{Fuzzy sphere harmonics}
In this appendix, we summarize the fuzzy sphere harmonics based on the
work \cite{istt}.
The fuzzy sphere harmonics is the eigen function on a
set of fuzzy $S^2$ with different radii.
Let us consider a set of linear maps ${\cal M}_{jj'}$ from a
$\left(2j+1\right)$-dimensional complex vector space $V_{j}$ to a
$\left(2j'+1\right)$-dimensional complex vector space $V_{j'}$, where
$j$ and $j'$ are non-negative half-integers.
${\cal M}_{jj'}$ is
$\left[\left(2j+1\right)\times\left(2j'+1\right)\right]$-dimensional
complex vector space.
They constructed a basis of ${\cal M}_{jj'}$ to use a basis of the spin
$j$ and $j'$ representations of $SU(2)$ as a basis of $V_j$ and
$V_{j'}$:
\begin{eqnarray}
 \left|jr\right>\!\!\left<j'r'\right|,
\end{eqnarray}
where $r=-j,-j+1,\cdots,j-1,j$ and $r'=-j',-j'+1,\cdots,j'-1,j'$.
Then, an arbitrary element of ${\cal M}_{jj'}$: $M$ is represented by
\begin{eqnarray}
 M=\sum_{rr'}M_{rr'}\left|jr\right>\!\!\left<j'r'\right|.
\end{eqnarray}
They defined linear maps from ${\cal M}_{jj'}$ to ${\cal M}_{jj'}$ by
its operation on the basis as follows:
\begin{eqnarray}
 L_i\circ\left|jr\right>\!\!\left<j'r'\right|
  =L_i\left|jr\right>\!\!\left<j'r'\right|
  -\left|jr\right>\!\!\left<j'r'\right|L_i,
\end{eqnarray}
where $L_i$ is a representation matrix (\ref{vacuum-pwmm}) for a $\hat{N}$-dimensional
representation of $SU(2)$.
The matrix element $M_{rr'}$ is transformed under linear maps from
${\cal M}_{jj'}$ to ${\cal M}_{jj'}$:
\begin{eqnarray}
 \left(L_i\circ M\right)_{rr'}
  =\left(L^{[j]}_i\right)_{rp}M_{pr'}
  -M_{rp'}\left(L^{[j']}_i\right)_{p'r'},
\end{eqnarray}
where $L^{[j]}_i$ is the $\left(2j+1\right)\times\left(2j+1\right)$
representation matrix of for the spin $j$ representation of $SU(2)$.
Additionally, the following identity holds:
\begin{eqnarray}
 \left(L_i\circ L_j\circ -L_j\circ L_i\circ\right)
  \left|jr\right>\!\!\left<j'r'\right|
  ={\rm i}\epsilon_{ijk}L^k\circ
  \left|jr\right>\!\!\left<j'r'\right|.
\end{eqnarray}
The fuzzy sphere harmonics is defined as a basis of ${\cal
M}_{jj'}$:
\begin{eqnarray}
 \hat{Y}_{JM}^{(jj')}
  =\sum_{rr'}
    \left(-1\right)^{3j-r'+2J+M-(\zeta+\zeta')/2}
    \sqrt{N_0\left(2J+1\right)}
    \left(
          \begin{array}{ccc}
           j  & J & j' \\
           -r & m & r' \\
          \end{array}
          \right)
    \left|jr\right>\!\!\left<j'r'\right|,
\end{eqnarray}
where $J=\left|j-j'\right|,\left|j-j'\right|+1,\cdots,j+j'-1,j+j'$
and $M=-J,-J+1,\cdots,J-1,J$. Furthermore, they have introduced a
positive integer $N_0$ as the following parameter:
\begin{eqnarray}
 2j+1=N_0+\zeta,
  \hspace{0.5cm}
 2j'+1=N_0+\zeta'.
\end{eqnarray}
Therefore,
$J=\left|\zeta-\zeta'\right|/2,\left|\zeta-\zeta'\right|/2+1,\cdots,\left(\zeta+\zeta'\right)/2+N_0-2,\left(\zeta+\zeta'\right)/2+N_0-1$.
$N_0$ plays a role of an ultraviolet cutoff scale for the angular
momentum. Since $\hat{Y}^{(jj')}_{JM}$ is the basis of the spin $J$
irreducible representation of $SU(2)$, the following equations hold
\begin{eqnarray}
 &&L_i\circ L^i\circ\hat{Y}^{(jj')}_{JM}
   =J\left(J+1\right)\hat{Y}^{(jj')}_{JM}, \nonumber \\
 &&L_3\circ\hat{Y}^{(jj')}_{JM}
   =M\hat{Y}^{(jj')}_{JM}.
\end{eqnarray}
$\hat{Y}^{(jj')}_{JM}$ is normalized as follows:
\begin{eqnarray}
 \frac{1}{N_0}{\rm Tr}\,\hat{Y}_{J_1M_1}^{(jj')}\hat{Y}_{J_2M_2}^{(j'j)}
  =\left(-1\right)^{M_1-(j-j')}\delta_{J_1J_2}\,\delta_{M_1-M_2},
\end{eqnarray}
where ${\rm Tr}$ stands for a trace over
$\left(2j+1\right)\times\left(2j+1\right)$ matrices. The hermitian
conjugate of $\hat{Y}_{JM}^{(jj')}$ is defined as
\begin{eqnarray}
 \hat{Y}_{JM}^{(jj')\dagger}=\left(-1\right)^{M-(j-j')}\hat{Y}_{J-M}^{(j'j)}.
\end{eqnarray}
The product of three fuzzy spherical harmonics can be evaluated as
\begin{eqnarray}
 \frac{1}{N_0}{\rm Tr}\,
  \hat{Y}_{J_1M_1}^{(jj')}\hat{Y}_{J_2M_2}^{(j'j'')}\hat{Y}_{J_3M_3}^{(j''j)}
  &=&\left(-1\right)^{-j+j'+J_1+J_2+J_3-\zeta/2-3\zeta'/2-\zeta''}\!\!
    \sqrt{N_0\left(2J_1+1\right)\left(2J_2+1\right)\left(2J_3+1\right)}
    \nonumber \\
 &&\times\left(
          \begin{array}{ccc}
           J_1 & J_2 & J_3 \\
           M_1 & M_2 & M_3 \\
          \end{array}
          \right)\!
    \left\{
          \begin{array}{ccc}
           J_1 & J_2 & J_3 \\
           j'' &  j  & j' \\
          \end{array}
          \right\},
\end{eqnarray}
where $(\cdots)$ and $\{\cdots\}$ represent the 3-$j$ and 6-$j$ symbol of
Wigner, respectively \cite{edmonds,vmk}.
In the large $N_0$ limit, we can obtain that
\begin{eqnarray}
 \frac{1}{N_0}{\rm Tr}\,
  \hat{Y}_{J_1M_1}^{(jj')}\hat{Y}_{J_2M_2}^{(j'j'')}\hat{Y}_{J_3M_3}^{(j''j)}
  &\longrightarrow&
  \left(-1\right)^{2J_2-2J_3-\tilde{M}_1}
   \sqrt{\left(2J_1+1\right)\left(2J_2+1\right)\left(2J_3+1\right)}
  \nonumber \\
 &&\times\left(
          \begin{array}{ccc}
           J_1 & J_2 & J_3 \\
           M_1 & M_2 & M_3 \\
          \end{array}
          \right)\!
    \left(
          \begin{array}{ccc}
                  J_1  &     J_2     &    J_3 \\
           \tilde{M_1} & \tilde{M_2} & \tilde{M_3} \\
          \end{array}
          \right),
    \label{3-fuzzy_harmonics}
\end{eqnarray}
where $\tilde{M}_1=j-j'$, $\tilde{M}_2=j'-j''$ and $\tilde{M}_3=j''-j$.
\section{Two-loop effective action on background $Y_i$}
In this appendix, we calculate the effective action at the 2-loop
level on the background $Y_i$. We can investigate the effective
action $\hat{W}$ of the deformed I\hspace{-.1em}IB matrix model on
the background $Y_i$ as follows:
\begin{eqnarray}
 \hat{W}&=&-\log\!\int\!da\,d\varphi\,dc\,db\;
    {\rm e}^{-\hat{S}} \nonumber \\
  &=&\hat{W}_{\rm tree}+\hat{W}_{\rm 1-loop}+\hat{W}_{\rm 2-loop},
\end{eqnarray}
where
\begin{eqnarray}
 \hat{W}_{\rm 2-loop}
  &=&\left<\exp\left\{{\rm Tr}\sum_{stu}
                \left([p_i, a^{(s,t)}_{\mu}][a^{(t,u)i},a^{(u,s)\mu}]
                 -b^{(s,t)}\Bigl[p_i,[a^{(t,u)i},c^{(u,s)}]\Bigr]\right.\right.\right.
               \nonumber \\
  &&\left.\left.\left.
           +\,\frac12\bar{\varphi}^{(s,t)}\Gamma^{\mu}[a^{(t,u)}_{\mu},\varphi^{(u,s)}]
             -\frac{{\rm i}}{3}f_{ijk}[a^{(s,t)i},a^{(t,u)j}]a^{(u,s)k}
             +\frac14\left[a^{(s,t)}_{\mu},a^{(t,u)}_{\nu}\right]^2
                \right)\right\}\right>_{\rm 1PI}, \nonumber \\
\end{eqnarray}
and $\left<\cdots\right>_{\rm 1PI}$ implies that we sum only the 1PI
diagrams.
The evaluation procedure
parallels to that of appendix A. There are precisely identical five
1PI diagrams to evaluate which are illustrated in Fig. 2.
\subsection{Bosonic propagators}
From the quadratic terms for the gauge fields $a^{(s,t)}_{\mu}$ in the
action (\ref{hat-s}), we can read out the propagators of gauge boson
modes $a^{(s,t)}_{\mu JM}$:
\begin{eqnarray}
 &&{\rm Tr}\sum_{st}
  \left(\frac12
   a^{(s,t)}_{\mu}
   P_i^2
   \delta^{\mu\nu}
   a^{(t,s)}_{\nu}\right)
  \nonumber \\
 &=&{\rm Tr}\sum_{st}
   \left(\frac12
    \sum_{J_1M_1}
    \sum_{J_2M_2}
   a^{(s,t)}_{\mu J_1M_1}\otimes
   \hat{Y}^{(j_sj_t)}_{J_1M_1}
   \beta^2J_2\left(J_2+1\right)
   \delta^{\mu\nu}
   a^{(t,s)}_{\nu J_2M_2}\otimes
   \hat{Y}^{(j_tj_s)}_{J_2M_2}\right)
    \nonumber \\
 &=&\frac12
    \sum_{st}
    \sum_{J_1M_1}
    \sum_{J_2M_2}
    a^{(s,t)}_{\mu J_1M_1}\,
    \beta^2J_2\left(J_2+1\right)
    \delta^{\mu\nu}
    (-1)^{-M_1+\left(j_s-j_t\right)}N_0\,
    \delta_{J_1J_2}\,
    \delta_{M_1-M_2}\,
    a^{(t,s)}_{\nu J_2M_2}.
    \nonumber \\
\end{eqnarray}
We expand the quantum fluctuations by the fuzzy spherical harmonics
as follows:
\begin{eqnarray}
 a^{(s,t)}_{\mu}=
  \sum_{JM}
  a^{(s,t)}_{\mu JM}\otimes
  \hat{Y}^{(j_sj_t)}_{JM}.
\end{eqnarray}
Therefore,
\begin{eqnarray}
 \left<a^{(s,t)}_{\mu J_1M_1}\,
  a^{(t,s)}_{\nu J_2M_2}\right>
  =\frac{1}{\beta^2N_0}\frac{(-1)^{M_1-(j_s-j_t)}}{J_1\left(J_1+1\right)}\,
   \delta_{\mu\nu}\,
   \delta_{J_1J_2}\,
   \delta_{M_1-M_2}.
\end{eqnarray}
The propagators of gauge boson fields become
\begin{eqnarray}
 \left<a^{(s,t)}_{\mu}\,a^{(t,s)}_{\nu}\right>
  =\frac{1}{\beta^2N_0}
   \sum_{JM}
   \frac{(-1)^{M-(j_s-j_t)}}{J\left(J+1\right)}\,
   \delta_{\mu\nu}\,
  \hat{Y}^{(j_sj_t)}_{JM}\,
  \hat{Y}^{(j_tj_s)}_{J-M}.
\end{eqnarray}
In the same way, we can read off the propagators of ghost fields:
\begin{eqnarray}
 &&\left<c^{(s,t)}_{J_1M_1}\,
    b^{(t,s)}_{J_2M_2}\right>
  =\frac{1}{\beta^2N_0}
   \frac{(-1)^{M_1-(j_s-j_t)}}{J_1\left(J_1+1\right)}\,
   \delta_{J_1J_2}\,
   \delta_{M_1-M_2},
    \nonumber \\
 &&\left<c^{(s,t)}\,b^{(t,s)}\right>
  =\frac{1}{\beta^2N_0}
   \sum_{JM}
   \frac{(-1)^{M-(j_s-j_t)}}{J\left(J+1\right)}\,
    \hat{Y}^{(j_sj_t)}_{JM}\,
    \hat{Y}^{(j_tj_s)}_{J-M}.
\end{eqnarray}
We expand the ghost fields by the fuzzy spherical harmonics as
follows:
\begin{eqnarray}
 &&c^{(s,t)}=
  \sum_{JM}c^{(s,t)}_{JM}\otimes
    \hat{Y}^{(j_sj_t)}_{JM}, \nonumber \\
 &&b^{(s,t)}=
  \sum_{JM}b^{(s,t)}_{JM}\otimes
    \hat{Y}^{(j_sj_t)}_{JM}.
\end{eqnarray}
\subsection{Contribution from four-point gauge boson vertex (a)}
We evaluate the 1PI diagram involving a 4-point gauge boson vertex:
\begin{eqnarray}
 \hat{V}_4&=&\frac14{\rm Tr}\sum_{stu}
       \left[a^{(s,t)}_{\mu},a^{(t,u)}_{\nu}\right]^2 \nonumber \\
    &=&\frac12{\rm Tr}\sum_{stuv}
       \left(a^{(s,t)}_{\mu}a^{(t,u)}_{\nu}a^{(u,v)\mu}a^{(v,s)\nu}
                      -a^{(s,t)}_{\mu}a^{(t,u)\mu}a^{(u,v)}_{\nu}a^{(v,s)\nu}\right).
\end{eqnarray}
We can calculate $\hat{V}_4$ by performing the Wick contraction.
\begin{eqnarray}
 \hat{V}_4
  &=&
   \left[\frac12\left(10+10\right)-\frac12\left(10^2+10\right)\right]
   {\rm Tr}\sum_{stuv}
   \sum_{J_1M_1}
   \sum_{J_2M_2}
   \left(\frac{1}{\beta^2N_0}\right)^2
   \frac{(-1)^{M_1+M_2-(j_s-j_t)-(j_u-j_v)}}
        {J_1\left(J_1+1\right)J_2\left(J_2+1\right)}
        \nonumber \\
  &&\times
   \hat{Y}^{(j_sj_t)}_{J_1M_1}\,
   \hat{Y}^{(j_tj_u)}_{J_1-M_1}\,
   \hat{Y}^{(j_uj_v)}_{J_2M_2}\,
   \hat{Y}^{(j_vj_s)}_{J_2-M_2}
   \nonumber \\
  &=&-45{\rm Tr}
   \sum_{stuv}
   \sum_{pq}
   \sum_{J_1M_1}
   \sum_{J_2M_2}
   \left(\frac{1}{\beta^2N_0}\right)^2
   \frac{(-1)^{M_1+M_2-(j_s-j_t)-(j_u-j_v)}}
        {J_1\left(J_1+1\right)J_2\left(J_2+1\right)}
        \nonumber \\
  &&\times
   \hat{Y}^{(j_sj_t)}_{J_1M_1}\,
   \hat{Y}^{(j_tj_p)}_{J_1-M_1}\,
   \hat{Y}^{(j_uj_v)}_{J_2M_2}\,
   \hat{Y}^{(j_vj_q)}_{J_2-M_2}\,
   \delta_{pu}
   \delta_{qs}.
\end{eqnarray}
Here we have inserted the complete set as follows:
\begin{eqnarray}
 \frac{1}{N_0}{\rm Tr}\sum_{J_3M_3}
  (-1)^{M_3-(j_p-j_q)}\,
  \hat{Y}^{(j_pj_u)}_{J_3M_3}\,
  \hat{Y}^{(j_qj_s)}_{J_3-M_3}
  =\delta_{pu}
   \delta_{qs}.
\end{eqnarray}
Therefore, we can get
\begin{eqnarray}
 \hat{V}_4
   =\frac{-45}{N_0}\,
    \hat{\sum_{123}}\,
    \hat{\Psi}_{123}^{\dagger}
    \frac{1}{\hat{P}^2\hat{Q}^2}
    \hat{\Psi}_{123},
\end{eqnarray}
where $\hat{P}$, $\hat{Q}$ and $\hat{R}$ are defined as follows:
\begin{eqnarray}
 &&\hat{P}_i\hat{Y}^{(j_sj_t)}_{J_1M_1}
  \equiv
  \left[p_i,\hat{Y}^{(j_sj_t)}_{J_1M_1}\right],
  \nonumber \\
 &&\hat{Q}_i\hat{Y}^{(j_sj_t)}_{J_2M_2}
  \equiv
  \left[p_i,\hat{Y}^{(j_sj_t)}_{J_2M_2}\right],
  \nonumber \\
 &&\hat{R}_i\hat{Y}^{(j_sj_t)}_{J_3M_3}
  \equiv
  \left[p_i,\hat{Y}^{(j_sj_t)}_{J_3M_3}\right].
\end{eqnarray}
We have introduced the following wave function:
\begin{eqnarray}
 \hat{\Psi}_{123}
  \equiv \frac{1}{N_0}
         {\rm Tr}\sum_{stu}
         \hat{Y}^{(j_sj_t)}_{J_1M_1}\,
         \hat{Y}^{(j_tj_u)}_{J_2M_2}\,
         \hat{Y}^{(j_uj_s)}_{J_3M_3}.
\end{eqnarray}
$\hat{\sum}_{123}$ denotes
$\sum_{J_1M_1}\sum_{J_2M_2}\sum_{J_3M_3}$.
\subsection{Contribution from three-point gauge boson vertex (b)}
We evaluate the 1PI diagram involving 3-point gauge boson
vertices. We can express the contribution corresponding to the
diagram (b) as follows:
\begin{eqnarray}
 \hat{V}_3=\frac12\left({\rm
                   Tr}\sum_{stu}\left[p_i,a^{(s,t)}_{\mu}\right]
                   \left[a^{(t,u)i},a^{(u,s)\mu}\right]\right)^2.
\end{eqnarray}
We can express the result as a following compact form:
\begin{eqnarray}
 \hat{V}_3
  =\frac{9}{2N_0}
  \hat{\sum_{123}}
  \hat{\Psi}_{123}^{\dagger}\,
  \frac{2\hat{P}^2-\hat{P}\cdot \hat{Q}-\hat{P}\cdot \hat{R}}
       {\hat{P}^2\hat{Q}^2\hat{R}^2}\,
  \hat{\Psi}_{123}.
\end{eqnarray}
We use the following relation:
\begin{eqnarray}
 \hat{P}\cdot \hat{Q}\,\hat{\Psi}_{123}
  =\frac{\hat{R}^2-\hat{P}^2-\hat{Q}^2}{2}\,
  \hat{\Psi}_{123},
\end{eqnarray}
and the momentum conserved relation: $\hat{P}+\hat{Q}+\hat{R}=0$.
Therefore, we can simplify the result as
\begin{eqnarray}
 \hat{V}_3
  =\frac{27}{3N_0}
   \hat{\sum_{123}}
   \hat{\Psi}_{123}^{\dagger}
   \frac{1}{\hat{P}^2\hat{Q}^2}
   \hat{\Psi}_{123}.
\end{eqnarray}
We note the following relation
\begin{eqnarray}
 \sum_{M}\,
  (-1)^{M-(j_s-j_t)}
  \hat{Y}^{(j_sj_t)}_{JM}
  \hat{P}_i\,
  \hat{Y}^{(j_tj_s)}_{J-M}
  =-\sum_{M}\,
  (-1)^{M-(j_s-j_t)}
  \left(\hat{P}_i\,\hat{Y}^{(j_sj_t)}_{JM}\right)
  \hat{Y}^{(j_tj_s)}_{J-M}.
\end{eqnarray}
\subsection{Contribution from the Myers type interaction (c)}
The diagram involving the Myers type interactions is represented as
follows:
\begin{eqnarray}
 \hat{V}_{\rm M}=
  -\frac{1}{18}
  \left({\rm Tr}\sum_{stu}f_{ijk}\left[a^{(s,t)i},a^{(t,u)j}\right]a^{(u,s)k}\right)^2.
\end{eqnarray}
We can evaluate it as
\begin{eqnarray}
 \hat{V}_{\rm M}
  =\frac{4\beta^2}{N_0}\,
   \hat{\sum_{123}}\,
   \hat{\Psi}_{123}^{\dagger}\,
   \frac{1}{\hat{P}^2\hat{Q}^2\hat{R}^2}\,
   \hat{\Psi}_{123},
\end{eqnarray}
where we have used the relation: $f_{ijk}f^{ijk}=6\beta^2$.
\subsection{Contribution from the ghost interaction (d)}
We evaluate the contribution from the ghost interactions.
\begin{eqnarray}
 \hat{V}_{\rm gh}
  =\frac12\left({\rm Tr}\sum_{stu}\left[p_i,b^{(s,t)}\right]
           \left[a^{(t,u)i},c^{(u,s)}\right]\right)^2.
\end{eqnarray}
The result is
\begin{eqnarray}
 \hat{V}_{\rm gh}
  =-\frac{1}{2N_0}
   \hat{\sum_{123}}\,
   \hat{\Psi}_{123}^{\dagger}\,
   \frac{1}{\hat{P}^2\hat{Q}^2}\,
   \hat{\Psi}_{123}.
\end{eqnarray}
\subsection{Fermion propagator}
We can read off the fermion propagator from the quadratic term of
$\varphi$ in the action (\ref{hat-s}):
\begin{eqnarray}
 &&{\rm Tr}\sum_{st}
  \left(-\frac12\,\bar{\varphi}^{(s,t)}\Gamma^iP_i\varphi^{(t,s)}\right)
   \nonumber \\
 &=&
   {\rm Tr}\sum_{st}
   \left(-\frac12
    \sum_{J_1M_1}
    \sum_{J_2M_2}
   \bar{\varphi}^{(s,t)}_{J_1M_1}\otimes
   \hat{Y}^{(j_sj_t)\dagger}_{J_1M_1}\,
   \Gamma^iP_i\,
   \varphi^{(t,s)}_{J_2M_2}\otimes
   \hat{Y}^{(j_tj_s)}_{J_2M_2}
  \right) \nonumber \\
 &=&-\frac12
   \sum_{st}
   \sum_{J_1M_1}
   \sum_{J_2M_2}
   \bar{\varphi}^{(s,t)}_{J_1M_1}\,
   \Gamma^iP_iN_0\,
   \delta_{J_1J_2}\,
   \delta_{M_1M_2}\,
   \varphi^{(t,s)}_{J_2M_2},
\end{eqnarray}
where we expanded the quantum fluctuations by the fuzzy spherical
harmonics as follows:
\begin{eqnarray}
 \varphi^{(s,t)}
  =\sum_{JM}
       \varphi^{(s,t)}_{JM}\otimes
       \hat{Y}^{(s,t)}_{JM}.
\end{eqnarray}
We thus find
\begin{eqnarray}
 \left<\varphi^{(s,t)}_{J_1M_1}\,
  \bar{\varphi}^{(t,s)}_{J_2M_2}\right>
 =-\frac{1}{N_0\Gamma^iP_i}\,
   \delta_{J_1J_2}\,
   \delta_{M_1M_2}.
\end{eqnarray}
The fermion propagator is
\begin{eqnarray}
 \left<\varphi^{(s,t)}\,
  \bar{\varphi}^{(t,s)}\right>
 =\sum_{JM}
  \left(-\frac{1}{N_0\Gamma^iP_i}\right)
  (-1)^{M-(j_s-j_t)}
  \hat{Y}^{(j_sj_t)}_{JM}
  \hat{Y}^{(j_tj_s)}_{J-M}.
\end{eqnarray}
We can further expand the fermion propagator in powers of $1/P^2$ as
in appendix A.6.
\begin{eqnarray}
 -\frac{1}{\Gamma^iP_i}
  =-\frac{1}{P^2}\,\Gamma^iP_i
   +\frac{{\rm i}}{2}\left(\frac{1}{P^2}\right)^2\!\!f_{ijk}\Gamma^{ijl}P^{k}P_{l}
   +{\cal O}\left(\left(\frac{1}{P^2}\right)^3\right).
\end{eqnarray}
\subsection{Contribution from the fermion interaction (e)}
Finally, we evaluate the diagram involving fermion interactions.
\begin{eqnarray}
 \hat{V}_{\rm F}
  &=&\frac18
  \left({\rm Tr}
   \sum_{stu}
   \bar{\varphi}^{(s,t)}
   \Gamma^{\mu}
   \left[a^{(t,u)}_{\mu},\varphi^{(u,s)}\right]\right)^2
      \nonumber \\
  &=&\frac12
   \left({\rm Tr}\sum_{stu}
    \bar{\varphi}^{(s,t)}
    \Gamma^{\mu}
    a^{(t,u)}_{\mu}
    \varphi^{(u,s)}\right)^2.
\end{eqnarray}
We can perform the Wick contractions:
\begin{eqnarray}
 \hat{V}_{\rm F}
  &=&\frac12\,{\rm Tr}\sum_{stu}
              {\rm Tr}\sum_{pqr}
  \nonumber \\
  &&\times
    \sum_{J_1M_1}
    \frac{1}{N_0}
    \left(\frac{1}{P^2}\,\Gamma^iP_i
     -\frac{{\rm i}}{2}\left(\frac{1}{P^2}\right)^2\!\!f_{ijk}\Gamma^{ijl}P^kP_l
     +\cdots\right)\!\!
     \left(\frac{1+\Gamma_{11}}{2}\right)\Gamma^{\mu}
     \nonumber \\
  &&\times (-1)^{M_1-(j_p-j_r)}
    \hat{Y}^{(j_sj_t)}_{J_1M_1}
    \hat{Y}^{(j_rj_p)}_{J_1-M_1}
    \nonumber \\
  &&\times\sum_{J_2M_2}
    \frac{1}{\beta^2N_0}
    \frac{(-1)^{M_2-(j_r-j_q)}}{J_2\left(J_2+1\right)}\,
    \delta_{\mu\nu}\,
    \hat{Y}^{(j_tj_u)}_{J_2M_2}\,
    \hat{Y}^{(j_qj_r)}_{J_2-M_2}
    \nonumber \\
  &&\times\sum_{J_3M_3}
    \frac{1}{N_0}
    \left(-\frac{1}{P^2}\,\Gamma^aP_a
    +\frac{{\rm i}}{2}\left(\frac{1}{P^2}\right)^2\!\!f_{abc}\Gamma^{abd}P^cP_d
    +\cdots\right)\!\!
    \left(\frac{1+\Gamma_{11}}{2}\right)\Gamma^{\nu}
    \nonumber \\
  &&\times (-1)^{M_3-(j_q-j_p)}
    \hat{Y}^{(j_uj_s)}_{J_3M_3}
    \hat{Y}^{(j_pj_q)}_{J_3-M_3}. \nonumber \\
\end{eqnarray}
We can evaluate the traces of products of gamma matrices as in
appendix A.7. We obtain the following result
\begin{eqnarray}
 \hat{V}_{\rm F}
  \sim
   \hat{\sum_{123}}
   \frac{1}{N_0}
   \hat{\Psi}_{123}^{\dagger}
   \left[-64\,\frac{\hat{P}\cdot \hat{Q}}{\hat{P}^2\hat{Q}^2\hat{R}^2}
    +32\beta^2\,\frac{\hat{P}^2\hat{Q}^2}
                     {(\hat{P}^2)^2(\hat{Q}^2)^2\hat{R}^2}
   \right]
   \hat{\Psi}_{123}.
   \nonumber \\
\end{eqnarray}
\subsection{Two-loop effective action}
We summarize the 2-loop effective action on the background $Y_i$
as follows:
\begin{eqnarray}
 \hat{W}_{\rm 2-loop}
  &\sim&
   \hat{V}_4+\hat{V}_3+\hat{V}_{\rm M}+\hat{V}_{\rm gh}+\hat{V}_{\rm F}
   \nonumber \\
  &=&\frac{36\beta^2}{N_0}
     \hat{\sum_{123}}\,
     \hat{\Psi}_{123}^{\dagger}\,
     \frac{1}{\hat{P}^2\hat{Q}^2\hat{R}^2}\,
     \hat{\Psi}_{123}
     \nonumber \\
  &\longrightarrow&
  \frac{36}{\beta^4N_0}
  \sum_{stu}
  \sum_{J_1J_2J_3}
   \frac{\left(2J_1+1\right)\left(2J_2+1\right)\left(2J_3+1\right)}
        {J_1\left(J_1+1\right)J_2\left(J_2+1\right)J_3\left(J_3+1\right)}
  \left(
   \begin{array}{ccc}
        J_1    &     J_2     &    J_3 \\
   \tilde{M}_1 & \tilde{M}_2 & \tilde{M}_3 \\
   \end{array}
         \right)^2.
  \nonumber \\
\end{eqnarray}
Here we have used the formula (\ref{3-fuzzy_harmonics}), and
performed the summations over $M_1$, $M_2$ and $M_3$.
\subsection{Two-loop effective action at large $N_0$ limit}
We evaluate the 2-loop effective action in such a large $N_0$
limit that $N_0\gg\Lambda$. We impose cutoff scale $2\Lambda$ on
$h$, and separate the summation over $J_1$, $J_2$ and $J_3$ into two
parts at a cutoff scale $\Lambda$ as (\ref{w-2loop-cutoff}). The
first term of (\ref{w-2loop-cutoff}) is calculated by using the
result of (\ref{sum-n1_n2_n3}):
\begin{eqnarray}
 \hat{W}_{\rm 2-loop}^{\,(1)}
 &\equiv&
  \frac{36}{\beta^4N_0}
  \sum_{h=1}^{2\Lambda}
  \sum_{J_1=1/2}^{\Lambda}
  \sum_{\tilde{M}_1=-J_1}^{J_1}
  \sum_{J_2=1/2}^{\Lambda}
  \sum_{\tilde{M}_2=-J_2}^{J_2}
  \sum_{J_3=1/2}^{\Lambda}
  \sum_{\tilde{M}_3=-J_3}^{J_3}
   \nonumber \\
  &&\times
   \frac{\left(2J_1+1\right)\left(2J_2+1\right)\left(2J_3+1\right)}
        {J_1\left(J_1+1\right)J_2\left(J_2+1\right)J_3\left(J_3+1\right)}
  \left(
   \begin{array}{ccc}
         J_1    &     J_2     &    J_3 \\
    \tilde{M}_1 & \tilde{M}_2 & \tilde{M}_3 \\
   \end{array}
         \right)^2
  \nonumber \\
 &=&
  \frac{36}{\beta^4N_0}
  \sum_{h=1}^{2\Lambda}
  \sum_{J_1=1/2}^{\Lambda}
  \sum_{J_2=1/2}^{\Lambda}
  \sum_{J_3=1/2}^{\Lambda}
   \frac{\left(2J_1+1\right)\left(2J_2+1\right)\left(2J_3+1\right)}
        {J_1\left(J_1+1\right)J_2\left(J_2+1\right)J_3\left(J_3+1\right)},
\end{eqnarray}
where we have evaluated the summations over $\tilde{M}_1$,
$\tilde{M}_2$ and $\tilde{M}_3$. We can assume that
$J_1,J_2,J_3\gg1$:
\begin{eqnarray}
 \hat{W}_{\rm 2-loop}^{\,(1)}
  &=&
  \frac{36}{\beta^4N_0}
  \sum_{h=1}^{2\Lambda}
  \sum_{J_1=1/2}^{\Lambda}
  \sum_{J_2=1/2}^{\Lambda}
  \sum_{J_3=1/2}^{\Lambda}
   \frac{\left(2J_1\right)\left(2J_2\right)\left(2J_3\right)}
        {J_1^2J_2^2J_3^2}
  \nonumber \\
 &\sim&
 \frac{2304}{\beta^4N_0}
 \sum_h
 \int_{1}^{2\Lambda}dn_1
 \int_{1}^{2\Lambda}dn_2
 \int_{1}^{2\Lambda}dn_3
 \frac{1}{n_1n_2n_3},
\end{eqnarray}
 where we have set that $J_1=n_1/2$, $J_2=n_2/2$ and $J_3=n_3/2$.
 $\left|n_2-n_3\right|\leq n_1\leq n_2+n_3$ and $n_1+n_2+n_3$ are
even numbers. We can obtain the following result by citing
(\ref{sum-n1_n2_n3}):
\begin{eqnarray}
 \hat{W}_{\rm 2-loop}^{\,(1)}
  \mathop{\longrightarrow}_{{N_0\rightarrow\infty}\atop{N_0\gg
  \Lambda}}
  \sum_h\left(\frac{576\pi^2}{\beta^4N_0}\log\Lambda\right).
\end{eqnarray}

The second term in (\ref{w-2loop-cutoff}) is calculated as follows:
\begin{eqnarray}
 \hat{W}_{\rm 2-loop}^{\,(2)}
  &\equiv&
   \frac{3\cdot36}{\beta^4N_0}
    \sum_{h=1}^{2\Lambda}
    \sum_{J_1=\Lambda+1/2}^{N_0}
    \sum_{J_2=1/2}^{\Lambda}
    \sum_{\tilde{M}_2=-J_2}^{J_2}
    \sum_{J_3=1/2}^{\Lambda}
    \sum_{\tilde{M}_3=-J_3}^{J_3}
   \nonumber \\
  &&\times\left(2\Lambda+1\right)
   \frac{\left(2J_1+1\right)\left(2J_2+1\right)\left(2J_3+1\right)}
  {J_1\left(J_1+1\right)J_2\left(J_2+1\right)J_3\left(J_3+1\right)}
  \left(
   \begin{array}{ccc}
         J_1    &     J_2     &    J_3 \\
        \Lambda & \tilde{M}_2 & \tilde{M}_3 \\
   \end{array}
         \right)^2
  \nonumber \\
  &\sim&
   \frac{3\cdot36}{\beta^4N_0}
    \sum_{h=1}^{2\Lambda}
    \sum_{J_1=\Lambda+1/2}^{N_0}
    \sum_{J_2=1/2}^{\Lambda}
    \sum_{\tilde{M}_2=-J_2}^{J_2}
    \sum_{J_3=1/2}^{\Lambda}
    \sum_{\tilde{M}_3=-J_3}^{J_3}
   \nonumber \\
  &&\times\left(2\Lambda+1\right)
   \frac{\left(2J_1+1\right)\left(2J_2+1\right)\left(2J_3+1\right)}
  {J_1\left(J_1+1\right)J_2\left(J_2+1\right)J_3\left(J_3+1\right)}
  \left(
   \begin{array}{ccc}
         J_1    &     J_2     &    J_3 \\
          0     & \tilde{M}_2 & \tilde{M}_3 \\
   \end{array}
         \right)^2
  \nonumber \\
  &=&
   \frac{3\cdot36}{\beta^4N_0}
    \sum_{h=1}^{2\Lambda}
    \sum_{J_1=\Lambda+1/2}^{N_0}
    \sum_{J_2=1/2}^{\Lambda}
    \sum_{J_3=1/2}^{\Lambda}
   \nonumber \\
  &&\times\left(2\Lambda+1\right)
   \frac{\left(2J_1+1\right)\left(2J_2+1\right)\left(2J_3+1\right)}
  {J_1\left(J_1+1\right)J_2\left(J_2+1\right)J_3\left(J_3+1\right)}
  \frac{1}{2J_1+1},
\end{eqnarray}
where we can assume that $J_1\sim N_0$ and $N_0\gg\Lambda$ in the
3-$j$ symbol, and perform the summations  over $\tilde{M}_2$ and
$\tilde{M}_3$. Additionally, we can assume that $J_1,J_2,J_3\gg1$:
\begin{eqnarray}
 \hat{W}_{\rm 2-loop}^{\,(2)}
  &\sim&
    \frac{3\cdot36}{\beta^4N_0}
    \sum_{h=1}^{2\Lambda}
    \sum_{J_1=\Lambda+1/2}^{N_0}
    \sum_{J_2=1/2}^{\Lambda}
    \sum_{J_3=1/2}^{\Lambda}
  \left(2\Lambda\right)
   \frac{\left(2J_1\right)\left(2J_2\right)\left(2J_3\right)}
  {J_1^2J_2^2J_3^2}
  \frac{1}{2J_1}
  \nonumber \\
  &\sim&
    \frac{13824\Lambda}{\beta^4N_0}
    \sum_h
    \int_{2\Lambda+1}^{2N_0}dn_1
    \int_{1}^{2\Lambda}dn_2
    \int_{1}^{2\Lambda}dn_3
    \frac{1}{n_1^2n_2n_3},
\end{eqnarray}
 where we have set that $J_1=n_1/2$, $J_2=n_2/2$ and $J_3=n_3/2$.
Therefore, we can estimate it as
\begin{eqnarray}
 \hat{W}_{\rm 2-loop}^{\,(2)}
  \mathop{\longrightarrow}_{{N_0\rightarrow\infty}\atop{N_0\gg\Lambda}}
  \sum_h
  \left({\rm const}\right),
\end{eqnarray}

The third term in (\ref{w-2loop-cutoff}) is calculated as follows:
\begin{eqnarray}
 \hat{W}_{\rm 2-loop}^{\,(3)}
  &\equiv&
   \frac{3\cdot36}{\beta^4N_0}
   \sum_{h=1}^{2\Lambda}
   \sum_{J_1=\Lambda+1/2}^{N_0}
   \sum_{J_2=\Lambda+1/2}^{N_0}
   \sum_{J_3=1/2}^{\Lambda}
   \sum_{\tilde{M}_3=-J_3}^{J_3}
   \nonumber \\
  &&\times
   \left(2\Lambda+1\right)^2
   \frac{\left(2J_1+1\right)\left(2J_2+1\right)\left(2J_3+1\right)}
        {J_1\left(J_1+1\right)J_2\left(J_2+1\right)J_3\left(J_3+1\right)}
  \left(
   \begin{array}{ccc}
         J_1    &     J_2     &    J_3 \\
      \Lambda   &   \Lambda   &  \tilde{M}_3  \\
   \end{array}
         \right)^2
  \nonumber \\
 &\sim&
  \frac{3\cdot36}{\beta^4N_0}
   \sum_{h=1}^{2\Lambda}
   \sum_{J_1=\Lambda+1/2}^{N_0}
   \sum_{J_2=\Lambda+1/2}^{N_0}
   \sum_{J_3=1/2}^{\Lambda}
   \sum_{\tilde{M}_3=-J_3}^{J_3}
   \nonumber \\
  &&\times
   \left(2\Lambda+1\right)^2
   \frac{\left(2J_1+1\right)\left(2J_2+1\right)\left(2J_3+1\right)}
        {J_1\left(J_1+1\right)J_2\left(J_2+1\right)J_3\left(J_3+1\right)}
  \left(
   \begin{array}{ccc}
         J_1    &     J_2     &    J_3 \\
          0     &      0      &  \tilde{M}_3  \\
   \end{array}
         \right)^2,
\end{eqnarray}
where we can assume the conditions that $J_1,J_2\sim N_0$ and
$N_0\gg\Lambda$ in the 3-$j$ symbol. Additionally, we can assume the
condition that $J_1,J_2,J_3\gg1$ and $J_1,J_2\gg J_3$:
\begin{eqnarray}
 \hat{W}_{\rm 2-loop}^{\,(3)}
  &\sim&
  \frac{3\cdot36}{\beta^4N_0}
   \sum_{h=1}^{2\Lambda}
   \sum_{J_1=\Lambda+1/2}^{N_0}
   \sum_{J_2=\Lambda+1/2}^{N_0}
   \sum_{J_3=1/2}^{\Lambda}
   \left(2\Lambda\right)^2
   \frac{\left(2J_1\right)\left(2J_2\right)\left(2J_3\right)}
        {J_1^2J_2^2J_3^2}
  \nonumber \\
 &&\times\frac{\left(-1\right)^{J_1-J_2+J_3}}{2J_2+1}
   \delta_{J_1-J_2+J_3,2n}
   \frac{\left(J_1-J_2+J_3\right)!\left(-J_1+J_2+J_3\right)!}
        {2^{2J_3}
        \left[\left(\frac{J_1-J_2+J_3}{2}\right)!
        \left(\frac{-J_1+J_2+J_3}{2}\right)!\right]^2},
\end{eqnarray}
where we have used the following relation for the squares of 3-$j$
symbol in the semi-classical limit \cite{vmk}:
\begin{eqnarray}
 \sum_{M_3}
  \left(
   \begin{array}{ccc}
         J_1    &     J_2     &    J_3 \\
          0     &      0      &     M_3  \\
   \end{array}
         \right)^2
  \sim
   \frac{1}{2J_2+1}
   \left[D_{0,-J_1+J_2}^{J_3}\left(0,\frac{\pi}{2},0\right)\right]^2.
\end{eqnarray}
The Wigner $D$-functions are given by
\begin{eqnarray}
 D_{0M}^{J}\left(0,\frac{\pi}{2},0\right)
  =\left(-1\right)^{\frac{J-M}{2}}
  \delta_{J-M,2n}
  \frac{\sqrt{\left(J-M\right)!\left(J+M\right)!}}
       {2^J\left(\frac{J+M}{2}\right)!\left(\frac{J-M}{2}\right)!},
\end{eqnarray}
where $\delta_{J-M,2n}$ implies that $J-M$ must be even numbers.
After making use of Stirling's formula, we obtain
\begin{eqnarray}
 \hat{W}_{\rm 2-loop}^{\,(3)}
  &\sim&
  \frac{3\cdot36}{\beta^4N_0}
   \sum_{h=1}^{2\Lambda}
   \sum_{J_1=\Lambda+1/2}^{N_0}
   \sum_{J_2=\Lambda+1/2}^{N_0}
   \sum_{J_3=1/2}^{\Lambda}
   \left(2\Lambda\right)^2
   \frac{\left(2J_1\right)\left(2J_2\right)\left(2J_3\right)}
        {J_1^2J_2^2J_3^2}
  \nonumber \\
 &&\times\frac{\left(-1\right)^{J_1-J_2+J_3}}{2J_2+1}
   \delta_{J_1-J_2+J_3,2n}
   \frac{2}{\pi}
   \frac{1}{\sqrt{\left(J_1-J_2+J_3\right)\left(-J_1+J_2+J_3\right)}}
   \nonumber \\
 &\sim&
  \frac{110592\Lambda^2}{\pi\beta^4N_0}
  \sum_h
  \int_{2\Lambda+1}^{2N_0}dn_1
  \int_{2\Lambda+1}^{2N_0}dn_2
  \int_{1}^{2\Lambda}dn_3
  \frac{1}{n_1n_2^2n_3}
  \frac{1}{\sqrt{n_3^2-\left(n_1-n_2\right)^2}},
  \nonumber \\
\end{eqnarray}
 where we have set that $J_1=n_1/2$, $J_2=n_2/2$ and $J_3=n_3/2$.
Therefore, we can estimate it as
\begin{eqnarray}
 \hat{W}_{\rm 2-loop}^{\,(3)}
  \mathop{\longrightarrow}_{{N_0\rightarrow\infty}\atop{N_0\gg\Lambda}}
  \sum_h
  \left({\rm const}\right),
\end{eqnarray}

The forth term in (\ref{w-2loop-cutoff}) is calculated as follows:
\begin{eqnarray}
 \hat{W}_{\rm 2-loop}^{\,(4)}
  &\equiv&
  \frac{36}{\beta^4N_0}
   \sum_{h=1}^{2\Lambda}
   \sum_{J_1=\Lambda+1/2}^{N_0}
   \sum_{J_2=\Lambda+1/2}^{N_0}
   \sum_{J_3=\Lambda+1/2}^{N_0}
   \nonumber \\
  &&\times
   \left(2\Lambda+1\right)^2
   \frac{\left(2J_1+1\right)\left(2J_2+1\right)\left(2J_3+1\right)}
        {J_1\left(J_1+1\right)J_2\left(J_2+1\right)J_3\left(J_3+1\right)}
  \left(
   \begin{array}{ccc}
         J_1    &     J_2     &    J_3 \\
        \Lambda &   \Lambda   &  \Lambda \\
   \end{array}
         \right)^2
  \nonumber \\
 &\sim&
  \frac{36}{\beta^4N_0}
   \sum_{h=1}^{2\Lambda}
   \sum_{J_1=\Lambda+1/2}^{N_0}
   \sum_{J_2=\Lambda+1/2}^{N_0}
   \sum_{J_3=\Lambda+1/2}^{N_0}
   \nonumber \\
  &&\times
   \left(2\Lambda+1\right)^2
   \frac{\left(2J_1+1\right)\left(2J_2+1\right)\left(2J_3+1\right)}
        {J_1\left(J_1+1\right)J_2\left(J_2+1\right)J_3\left(J_3+1\right)}
  \left(
   \begin{array}{ccc}
         J_1    &     J_2     &    J_3 \\
          0     &      0      &     0 \\
   \end{array}
         \right)^2
\end{eqnarray}
where we assume the conditions that $J_1,J_2,J_3\sim N_0$ and
$N_0\gg\Lambda$ in the 3-$j$ symbol. Additionally, we assume the
condition that $J_1,J_2,J_3\gg1$:
\begin{eqnarray}
 \hat{W}_{\rm 2-loop}^{\,(4)}
  &\sim&
  \frac{36}{\beta^4N_0}
   \sum_{h=1}^{2\Lambda}
   \sum_{J_1=\Lambda+1/2}^{N_0}
   \sum_{J_2=\Lambda+1/2}^{N_0}
   \sum_{J_3=\Lambda+1/2}^{N_0}
   \left(2\Lambda\right)^2
   \frac{\left(2J_1\right)\left(2J_2\right)\left(2J_3\right)}
        {J_1^2J_2^2J_3^2}
   \nonumber \\
 &&\times
        \frac{1}{\pi}
        \frac{1}{\sqrt{-J_1^4-J_2^4-J_3^4+2J_1^2J_2^2+2J_2^2J_3^2+2J_3^2J_1^2}},
\end{eqnarray}
where we can make use of the following relation for the squares of
3-$j$ symbol in the semi-classical limit \cite{vmk}:
\begin{eqnarray}
 4\pi\left(
   \begin{array}{ccc}
          J_1 & J_2 & J_3 \\
          M_1 & M_2 & M_3  \\
   \end{array}
         \right)^2
 \sim
 \frac{\delta_{M_1+M_2+M_3,0}}
      {\sqrt{A^2+\frac14\left(J_1^2M_2M_3
                 +J_2^2M_3M_1
                 +J_3^2M_1M_2\right)}}
 \label{3j-classical}
\end{eqnarray}
and
\begin{eqnarray}
 16A^2=-J_1^4-J_2^4-J_3^4+2J_1^2J_2^2+2J_2^2J_3^2+2J_3^2J_1^2.
\end{eqnarray}
We obtain
\begin{eqnarray}
 \hat{W}_{\rm 2-loop}^{\,(4)}
  &\sim&
  \frac{36864\Lambda^2}{\pi\beta^4N_0}
  \sum_h
  \int_{2\Lambda+1}^{2N_0}dn_1
  \int_{2\Lambda+1}^{2N_0}dn_2
  \int_{2\Lambda+1}^{2N_0}dn_3
  \nonumber \\
  &&\times
   \frac{1}{n_1n_2n_3}
   \frac{1}{\sqrt{-n_1^4-n_2^4-n_3^4+2n_1^2n_2^2+2n_2^2n_3^2+2n_3^2n_1^2}},
\end{eqnarray}
 where we have set that $J_1=n_1/2$, $J_2=n_2/2$ and $J_3=n_3/2$.
Therefore, we can estimate it as
\begin{eqnarray}
 \hat{W}_{\rm 2-loop}^{\,(4)}
  \mathop{\longrightarrow}_{{N_0\rightarrow\infty}\atop{N_0\gg\Lambda}}
  \sum_h
  \left({\rm const}\right),
\end{eqnarray}

We conclude that the 2-loop effective action is given as follows
in such a large $N_0$ limit that $N_0\gg\Lambda$:
\begin{eqnarray}
 \hat{W}_{\rm 2-loop}
  \mathop{\longrightarrow}_{{N_0\rightarrow\infty}\atop{N_0\gg\Lambda}}
  \sum_h
  \left(\frac{576\pi^2}{\beta^4N_0}\log\Lambda\right).
\end{eqnarray}
%
%\newpage
%%%%%%%%%%%%%%%%%%%%%%%%%%%%%%%%%%%%%%%%%%%%%%%%%%%%%%%%%%%%%%%%%%%%%%%%%%%%%%%
%%%%%%%%%%%%%%%%%%%%%%%%%%%%%%%%%%% references %%%%%%%%%%%%%%%%%%%%%%%%%%%%%%%%
%%%%%%%%%%%%%%%%%%%%%%%%%%%%%%%%%%%%%%%%%%%%%%%%%%%%%%%%%%%%%%%%%%%%%%%%%%%%%%%

%
\end{document}